\documentclass[aps,prd,nofootinbib,twocolumn,amssymb,eqsecnum,notitlepage]{revtex4-1}

\usepackage{bm}
\usepackage{amsmath,amsthm,amssymb}
\usepackage{color}
\usepackage[dvipdfmx]{graphicx}

\allowdisplaybreaks[1]

\def\be{\begin{equation}}
\def\ee{\end{equation}}
\def\ba{\begin{eqnarray}}
\def\ea{\end{eqnarray}}
\def\l{\left}
\def\r{\right}
\def\f{\frac}

\def\omb{\Omega_b}
\def\omc{\Omega_c}
\def\omr{\Omega_r}

\begin{document}

\title{Most general cubic-order Horndeski Lagrangian allowing for scaling solutions \\
and the application to dark energy}

\author{Noemi Frusciante$^1$, Ryotaro Kase$^2$, 
Nelson J.~Nunes$^1$, and Shinji Tsujikawa$^2$ }

\smallskip
\affiliation{$^1$Instituto de Astrof\'isica e Ci\^encias do Espa\c{c}o, 
Faculdade de Ci\^encias da Universidade de Lisboa,  Campo Grande, PT1749-016 Lisboa, Portugal\\
\smallskip
$^2$Department of Physics, Faculty of Science, Tokyo University of Science,
1-3, Kagurazaka, Shinjuku-ku, Tokyo 162-8601, Japan}

\begin{abstract}

In cubic-order Horndeski theories where a scalar field $\phi$ is 
coupled to nonrelativistic matter with a field-dependent coupling 
$Q(\phi)$, we derive the most general Lagrangian having scaling 
solutions on the isotropic and homogenous cosmological background. 
For constant $Q$ including the case of vanishing coupling, 
the corresponding Lagrangian reduces to the form 
$L=Xg_2(Y)-g_3(Y)\square \phi$, where 
$X=-\partial_{\mu}\phi\partial^{\mu}\phi/2$ and $g_2, g_3$ 
are arbitrary functions of $Y=Xe^{\lambda \phi}$ 
with constant $\lambda$.
We obtain the fixed points of the scaling Lagrangian for constant $Q$
and show that the $\phi$-matter-dominated-epoch ($\phi$MDE)
is present for the cubic coupling 
$g_3(Y)$ containing inverse power-law functions of $Y$. 
The stability analysis around the fixed points indicates 
that the $\phi$MDE can be followed by a stable critical 
point responsible for the cosmic acceleration. 
We propose a concrete dark energy model allowing for such 
a cosmological sequence and show that the ghost and 
Laplacian instabilities can be avoided even 
in the presence of the cubic coupling.

\end{abstract}


\maketitle

%%%%%%%%%%%%%%%%%%%%%%%%
\section{Introduction}
%%%%%%%%%%%%%%%%%%%%%%%%

Two decades have passed since the discovery of the late-time 
cosmic acceleration \cite{SN1,SN2}, but the origin of this phenomenon 
is still unknown. The simplest candidate for dark energy is 
the cosmological constant $\Lambda$ \cite{Weinberg}, but it is still a challenging 
problem to relate the vacuum energy arising from particle physics 
with the observed dark energy scale. 
In the $\Lambda$-Cold-Dark-Matter ($\Lambda$CDM) model, there have
been also tensions between the values of the Hubble constant $H_0$ constrained from 
the Cosmic Microwave Background (CMB) \cite{Planck2015,Planck2018} 
and low-redshift measurements \cite{RiessH0}.
It is worthwhile to pursue alternative possibilities for 
realizing the cosmic acceleration and study whether they 
can be better fitted with observational data over the $\Lambda$CDM model.

A scalar field $\phi$ with an associated potential energy $V(\phi)$, 
dubbed quintessence, is one of the candidates for 
dark energy \cite{quin1,quin2,quin3,quin4,quin5,quin6}. 
For example, the late-time cosmic acceleration can be driven 
by runaway potentials like the exponential potential 
$V(\phi)=V_0 e^{-\lambda \phi}$ \cite{quin3,Pedro1,Pedro2,CLW}  
and the inverse power-law 
potential $V(\phi)=V_0 \phi^{-p}$ ($p>0$) \cite{Paul1,Paul2}. 
Since the potential energy increases toward the past, 
quintessence can alleviate the small energy-scale problem  
of the cosmological constant.

In particular, if the field density $\rho_{\phi}$ scales 
in the same manner as the background matter density 
$\rho_{m}$, such a scalar field can be compatible with 
the energy scale related to particle physics.
The cosmological solution along which the ratio 
$\rho_{\phi}/\rho_{m}$ remains constant is known 
as a scaling solution \cite{Pedro1,Pedro2,CLW,Liddle,Sahni,Skordis,Dodelson,Uzan,Bacci,Pet,Ohashi09,Chiba14,Amen14,Baha}. 
In quintessence, the exponential potential 
$V(\phi)=V_0 e^{-\lambda \phi}$ gives rise to the scaling 
solution for $\lambda^2>3(1+w_m)$, 
where $w_m$ is the matter equation of state. 
This solution can be responsible for the scaling radiation and matter 
eras. Since the cosmic acceleration occurs for $\lambda^2<2$, 
we require the modification of the 
potential at late time to exit from the scaling matter era.
For instance, this is possible by taking into account 
another shallow exponential 
potential \cite{Barreiro,Guo1,Guo2,Nunes:2003ff}.

In the presence of a direct coupling $Q(\phi)$ between 
the scalar field $\phi$ and nonrelativistic matter, there exists 
another type of scaling solutions called the 
$\phi$-matter-dominated-epoch ($\phi$MDE) 
for quintessence with the exponential 
potential \cite{Amenco1,Amenco2}. 
Extending the analysis to k-essence \cite{kinf,kes1,kes2} 
for constant $Q$, 
it was shown that the Lagrangian with scaling solutions 
is restricted to be of the form $L=X g_2 (Y)$, where 
$X=-\partial_{\mu}\phi\partial^{\mu}\phi/2$ and $g_2$ 
is an arbitrary function of 
$Y=Xe^{\lambda \phi}$ \cite{Piazza,Tsuji04,Tsuji06}. 
The derivation of the scaling k-essence Lagrangian is 
also possible for the field-dependent coupling 
$Q(\phi)$ \cite{Amendola06}.

Quintessence and k-essence belong to subclasses of 
most general 
scalar-tensor theories with second-order equations of 
motion--dubbed Horndeski theories \cite{Horndeski}. 
If we apply Horndeski theories to dark energy and impose the 
condition that the speed of gravitational waves on the 
cosmological background is equivalent to that of 
light (as constrained from the GW170817 
event \cite{GW170817} together 
with the electromagnetic counterpart \cite{Goldstein}), 
the Lagrangian is of the form 
$L=G_2(\phi,X)-G_3(\phi,X) \square \phi+G_4(\phi)R$, 
where $G_4$ depends on $\phi$ alone with the Ricci 
scalar $R$, and $G_2, G_3$ are functions of 
$\phi$ and 
$X$ \cite{Lon15,GWcon1,GWcon2,GWcon3,GWcon4,GWcon5,GWcon6}.

In cubic-order Horndeski theories with $G_4={\rm constant}$, 
the Lagrangian with scaling solutions was derived 
in Ref.~\cite{Gomes1} for the $\phi$-dependent 
coupling $Q(\phi)$ (see also Ref.~\cite{Gomes2}). 
They imposed a particular ansatz (Eq.~(4.5) of Ref.~\cite{Gomes1}) 
for deriving the scaling Lagrangian, in addition to the choice of a 
specific form of the coupling $Q(\phi)=1/(c_1 \phi+c_2)$.
For constant $Q$, this led to the scaling 
Lagrangian of the form 
$L=Xg_2(Y)-g_3(Y)\square \phi$ with 
$g_3(Y)=a_1 Y+a_2 Y^2$, where 
$a_1$ and $a_2$ are constants.

On the other hand, the recent study of Ref.~\cite{Alb} showed 
that there exists a scaling solution for the cubic coupling 
$g_3(Y)=A \ln Y$, where $A$ is a constant, anticipating 
 the fact that the scaling solution may 
be present for a more general cubic coupling than that  
derived in Ref.~\cite{Gomes1}.
Indeed, for constant couplings $Q$ and $G_4$, it has been 
found that the cubic Lagrangian $-g_3(Y)\square \phi$ 
with an arbitrary function $g_3(Y)$ can allow for the 
existence of scaling solutions~\cite{Amen18}.

In this paper, we derive the most general cubic-order Horndeski 
Lagrangian with scaling solutions for a field-dependent coupling 
$Q(\phi)$. 
We show that, in the presence of the cubic 
Lagrangian $-G_3(\phi,X) \square \phi$, the coupling is constrained 
to be of the form $Q(\phi)=1/(c_1 \phi+c_2)$ for the existence of scaling 
solutions. 
For constant $Q$, the 
scaling Lagrangian reduces to the form 
$L=Xg_2(Y)-g_3(Y)\square \phi$, which is in agreement 
with the result of Ref.~\cite{Amen18}.
Moreover, our analysis encompasses the vanishing coupling 
($Q=0$) as a special case.

In the presence of the nonvanishing coupling constant $Q$, 
there exists a $\phi$MDE for the models in which the functions $g_2(Y)$ and  
$dg_3 (Y)/dY$ contain inverse powers $Y^{-n}$ ($n>0$), 
e.g., $g_2(Y)=c_0+c_1/Y$ and $g_3(Y)=d_1 \ln Y-d_2/Y$, 
where $c_0, c_1, d_1, d_2$ are constants.
The $\phi$MDE is characterized by the scaling solution with the  
field density parameter $\Omega_{\phi}$ affected by 
$Q$ and $d_1$.
This can lead to interesting 
cosmological solutions with the scaling saddle matter era 
followed by the late-time cosmic acceleration. 
The analysis in Ref.~\cite{Gomes1} overlooked 
the presence of $\phi$MDE in cubic Horndeski theories, as it is 
not present for the function $g_3(Y)=a_1 Y+a_2 Y^2$, while in Ref.~\cite{Amen18}, the 
authors did not consider a concrete cubic coupling $g_3(Y)$ with the $\phi$MDE.

We will show that, for the nonvanishing constant 
$Q$, there exist viable dark energy models with the $\phi$MDE in 
cubic-order Horndeski theories. We study the background cosmological 
dynamics by paying particular attention to the evolution of 
the field density parameter $\Omega_{\phi}$ and the dark energy 
equation of state $w_{\phi}$. Unlike the $\Lambda$CDM model, 
$\Omega_{\phi}$ does not need to be very much smaller than 
the background density parameters at early time.
Moreover, it is possible to avoid the ghost and Laplacian instabilities 
during the cosmic expansion history from the radiation era to today. 
Unlike the analysis of Ref.~\cite{Alb}, 
the Lagrangian does not need to be modified at late time to 
give rise to the cosmic acceleration. 

This paper is organized as follows.
In Sec.~\ref{Sec:back}, we formulate the coupled dark energy 
scenario in cubic-order Horndeski theories in terms of the 
Schutz-Sorkin action. 
In Sec.~\ref{Sec:derivelag}, the Lagrangian allowing 
for scaling solutions is generally derived for
the field-dependent matter coupling $Q(\phi)$. 
In Sec.~\ref{Sec:fixed}, we obtain the fixed points for the 
scaling Lagrangian with constant $Q$ and show the existence 
of $\phi$MDE for particular choices of $g_3(Y)$. 
In Sec.~\ref{Sec:sta}, we discuss the stability of the fixed points 
in the presence of nonrelativistic matter ($w_m=0$). 
In Sec.~\ref{Sec:dark}, we propose a concrete model of 
dark energy and study the dynamics of late-time cosmic 
acceleration preceded by the $\phi$MDE.
We conclude in Sec.~\ref{Sec:conclude}. 

Throughout the paper, we use the units where the 
speed of light $c$, the reduced Planck constant $\hbar$, 
and the reduced Planck mass $M_{\rm pl}$ are equivalent to 1.

%%%%%%%%%%%%%%%%%%%%%%%%%%%%%
\section{Background equations in cubic Horndeski theories}
\label{Sec:back}
%%%%%%%%%%%%%%%%%%%%%%%%%%%%%%

Let us consider the cubic-order Horndeski theories 
given by the action \cite{braiding}
\be
{\cal S} = 
\int d^{4}x \sqrt{-g}\l(\f{R}{2}
+L\r)+{\cal S}_m \left( \phi, g_{\mu \nu} \right)\,,
\label{action}
\ee
where $g$ is the determinant of metric tensor $g_{\mu \nu}$, 
$R$ is the Ricci scalar, and 
\be
L=G_2(\phi,X)-G_3(\phi,X) \square \phi\,.
\label{Lag}
\ee
The functions $G_2$ and $G_3$ depend on 
$\phi$ and $X=-\partial_{\mu}\phi\partial^{\mu}\phi/2$.
The Lagrangian (\ref{Lag}) belongs to the subclass of 
Horndeski theories with second-order equations 
of motion \cite{Horndeski,Horn1,Horn2,Horn3}. 
For this cubic-order Horndeski theory, 
the speed of gravitational waves $c_t$ on the cosmological background 
is equivalent to that of light \cite{Horn2,DT12}, so the model is consistent 
with the bound on $c_t$ constrained from the GW170817 event \cite{GW170817}.

For the matter action ${\cal S}_m$, we consider a barotropic 
perfect fluid coupled to the scalar field $\phi$.
Such a coupled system of matter and $\phi$ can be described 
by a Schutz-Sorkin action \cite{Sorkin,DGS,GPcosmo,GPGeff} 
with the matter density $\rho_m$ coupled to $\phi$.
In this case, the Schutz-Sorkin action is given by 
\be
{\cal S}_m=-\int d^4 x \left[ \sqrt{-g}\,\rho_{m} 
\left( n, \phi \right)+J^{\mu} \nabla_{\mu} \ell \right]\,,
\label{Sm}
\ee
where $\rho_m$ depends on the fluid 
number density $n$ as well as on $\phi$, and 
$\ell$ is a scalar quantity with the covariant derivative 
operator $\nabla_{\mu}$, and $J^{\mu}$ is 
a four vector related to $n$, as
\be
n=\sqrt{\frac{J^{\mu}J^{\nu} g_{\mu \nu}}{g}}\,.
\label{ndef}
\ee
Since we are now considering scalar-tensor theories, 
we do not need to take into account vector degrees of 
freedom in the action (\ref{Sm}).

We derive the background equations of motion on the flat 
Friedmann-Lema\^{i}tre-Robertson-Walker 
(FLRW) spacetime 
given by the line element 
\be
ds^2=-N^2(t)dt^2+a^2(t) \delta_{ij}dx^i dx^j\,,
\ee
where $N(t)$ is a lapse, and $a(t)$ is a scale factor. 
{}From Eq.~(\ref{ndef}), the background fluid number 
density $n_0$ is related to $J^0$, as $J^0=n_0 a^3$. 
Then, the matter action (\ref{Sm}) reduces to
\be
{\cal S}_m=-\int d^4x\,a^3 \left[ N \rho_m (n_0, \phi)
+n_0 \dot{\ell} \right]\,,
\label{Sm2}
\ee
where a dot represents a derivative with respect to 
the cosmic time $t$.
Varying the action (\ref{Sm2}) with respect to $\ell$, 
we find that  $n_0 a^3={\rm constant}$, i.e., 
\be
\dot{n}_0+3H n_0=0\,,
\label{n0eq}
\ee
where $H=\dot{a}/a$ is the Hubble expansion rate. 
This relation corresponds to the conservation 
of total fluid number. 

We vary the total action (\ref{action}) with respect to 
$N$ and $a$, and finally set $N=1$. 
This process leads to the following equations of motion:
\ba
& & 3H^2=\rho_{\phi}+\rho_m\,,\label{back1} \\
& & 2\dot{H}=-\rho_{\phi}-P_{\phi} 
-\rho_m -P_m\,,\label{back2}
\ea
where 
\ba
\hspace{-0.3cm}
\rho_{\phi}
&=&\dot{\phi}^2 G_{2,X}-G_2
-\dot{\phi}^2 \left( G_{3,\phi} 
-3H \dot{\phi} G_{3,X} \right),
\label{rhophi}\\
\hspace{-0.3cm}
P_{\phi}
&=& G_2-\dot{\phi}^2 \left( G_{3,\phi}
+\ddot{\phi}\,G_{3,X} \right)\,,
\label{Pphi}\\
\hspace{-0.3cm}
P_m &=& -n_0 \dot{\ell}-\rho_m\,,
\label{Pm}
\ea
with the notations $G_{i,\phi}=\partial G_i/\partial \phi$ 
and $G_{i,X}=\partial G_i/\partial X$. 
We note that $\rho_{\phi}$ and $P_{\phi}$ correspond to 
the field density and pressure, respectively.
Variation of the matter action with respect to $n_0$ leads 
to $\dot{\ell}=-\rho_{m,n}$, where 
$\rho_{m,n}=\partial \rho_m/\partial n_0$. 
Then, the matter pressure (\ref{Pm}) is expressed as 
\be
P_m=n_0 \rho_{m,n}-\rho_m\,.
\label{Pm2}
\ee
Taking the time derivative of $\rho_m=\rho_m (n_0, \phi)$, 
we obtain 
\be
\dot{\rho}_m=\rho_{m,n} \dot{n}_0
+Q(\phi) \rho_m \dot{\phi}\,,
\label{dotrhom}
\ee
where 
\be
Q(\phi) \equiv \frac{\rho_{m,\phi}}{\rho_m}\,.
\ee
Substituting Eq.~(\ref{n0eq}) into Eq.~(\ref{dotrhom}) and using 
Eq.~(\ref{Pm2}), it follows that 
\be
\dot{\rho}_{m}+3H \left( 1+w_{m}
\right) \rho_{m}=Q(\phi) \rho_m \dot{\phi}\,,
\label{back4}
\ee
where $w_m=P_m/\rho_m$.
Varying the action (\ref{action}) with respect to $\phi$, 
the scalar field obeys 
\be
\dot{\rho}_{\phi}+3H \left( 1+w_{\phi}
\right)\rho_{\phi}=-Q(\phi) \rho_m \dot{\phi}\,,
\label{back3}
\ee
where $w_{\phi}=P_{\phi}/\rho_{\phi}$.
The coupling $Q(\phi)$ quantifies the interaction 
between matter and $\phi$. 
For $Q(\phi)={\rm constant}$, the background 
matter density can be expressed in the form 
$\rho_m=\tilde{\rho}_m (n_0)e^{Q \phi}$, 
where $\tilde{\rho}_m$ is a function of $n_0$.

From Eq.~(\ref{back1}), the density parameters 
$\Omega_{\phi}=\rho_{\phi}/(3H^2)$ and 
$\Omega_{m}=\rho_{m}/(3H^2)$ obey the relation 
\be
\Omega_{\phi}+\Omega_{m}=1\,.
\label{Omepm}
\ee
{}From Eq.~(\ref{back2}), we have 
\be
\frac{\dot{H}}{H^2}=-\frac{3}{2} 
\left( 1+w_{\rm eff} \right)\,,\qquad 
w_{\rm eff}=w_{\phi}\Omega_{\phi}
+w_{m}\Omega_{m}\,,
\label{weff}
\ee
being $w_{\rm eff}$ the effective equation of state.

Solving Eqs.~(\ref{back2}) and (\ref{back3}) for 
$\ddot{\phi}$ and $\dot{H}$, we obtain
\begin{widetext}
\ba
\ddot{\phi} &=& [9G_{3,X}^2 H \dot{\phi}^5
+3G_{3,X} (G_{2,X}-2G_{3,\phi}) \dot{\phi}^4
-6G_{3, X\phi} H \dot{\phi}^3
+2 (G_{3,\phi \phi}-G_{2, X\phi}-9 G_{3,X}H^2) 
\dot{\phi}^2 \nonumber\\
& &
-6H (G_{2,X}-2G_{3,\phi}) \dot{\phi} 
+2G_{2,\phi}+3(\rho_m+P_m)G_{3,X} 
\dot{\phi}^2-2Q \rho_m]/q_s\,,\label{ddotphi}\\
\dot{H} &=&
[ -9G_{3,X} G_{3,XX}H^2 \dot{\phi}^6 
-3\{ G_{2,XX}G_{3,X}+G_{3,XX} (G_{2,X}-2G_{3,\phi})\}
H \dot{\phi}^5+\{ (G_{2,X}-2G_{3,\phi}) (G_{3,X \phi}
-G_{2,XX}) \nonumber \\
& &+G_{3,X} (G_{3,\phi \phi}-G_{2, X \phi}) 
-27 H^2 G_{3,X}^2 \} \dot{\phi}^4 
-12(G_{2,X}-2G_{3,\phi})G_{3,X} H \dot{\phi}^3
+(G_{2,\phi}G_{3,X}+4G_{2,X}G_{3,\phi} \nonumber \\
& &
-G_{2,X}^2-4G_{3,\phi}^2) \dot{\phi}^2 
-(\rho_m+P_m)(G_{2,X}+G_{2,XX}\dot{\phi}^2
-2G_{3,\phi}+6G_{3,X}H \dot{\phi} 
-G_{3,X \phi} \dot{\phi}^2+3G_{3,XX}H \dot{\phi}^3) 
\nonumber \\
& &
-Q \rho_m G_{3,X} \dot{\phi}^2]/q_s\,,
\label{tH}
\ea
where 
\be
q_s \equiv 3G_{3,X}^2 \dot{\phi}^4+6 G_{3,XX}  
H \dot{\phi}^3+2(G_{2,XX}-G_{3,X\phi}) \dot{\phi}^2
+12 G_{3,X} H \dot{\phi}
+2 (G_{2,X}-2G_{3,\phi})\,.
\label{qs}
\ee
\end{widetext}
The background dynamics is known by solving 
Eqs.~(\ref{back4}), (\ref{ddotphi}) and (\ref{tH})
together with the constraint Eq.~(\ref{back1}).

%%%%%%%%%%%%%%%%%%%%%%%%%%%%%
\section{Lagrangian allowing for scaling solutions}
\label{Sec:derivelag}
%%%%%%%%%%%%%%%%%%%%%%%%%%%%%

The scaling solution is characterized by a nonvanishing 
constant ratio $\Omega_{\phi}/\Omega_{m}$, so 
that both $\Omega_{\phi}$ and $\Omega_m$ are 
constant from Eq.~(\ref{Omepm}). 
Moreover, we would like to consider the case in which 
the field equation of state $w_{\phi}=P_{\phi}/\rho_{\phi}$ 
as well as $w_m$ do not vary in time in the scaling regime. 
Then, from Eq.~(\ref{weff}), both $w_{\rm eff}$ and 
$\dot{H}/H^2$ are constant.

In Sec.~\ref{scaderive1}, we first obtain the Lagrangian 
with scaling solutions for constant $Q$ and show that this 
agrees with the result recently found  in Ref.~\cite{Amen18}. 
This analysis also accommodates 
the vanishing coupling ($Q=0$) as a special case. 
In Sec.~\ref{scaderive2}, we derive the scaling Lagrangian 
for more general cases in which the coupling $Q$ 
depends on $\phi$.

\subsection{Constant $Q$ (including $Q=0$)}
\label{scaderive1}

For scaling solutions, both $\rho_{\phi}$ and $\rho_{m}$ 
are in proportion to $H^2$. 
Then, all the terms on the left hand side of 
Eq.~(\ref{back3}) are in proportion to $H^3$. 
For constant $Q$, the compatibility with the right hand 
side of Eq.~(\ref{back3}) shows that 
$\dot{\phi} \propto H$, i.e., 
\be
\frac{\dot{\phi}}{H}=\alpha\,,
\label{dphiH}
\ee
where $\alpha$ is a constant. 
The relation (\ref{dphiH}) is also consistent 
with Eq.~(\ref{back4}).
On using the scaling relation 
$\dot{\rho}_{\phi}/\rho_{\phi}=\dot{\rho}_m/\rho_m$ 
for $Q \neq 0$, it follows that  
$\alpha=3\Omega_{\phi}(w_m-w_{\phi})/Q$.

While Eq.~(\ref{dphiH}) has been derived for 
the nonvanishing constant $Q$, there are scaling 
solutions satisfying the condition (\ref{dphiH}) 
even for $Q=0$. For example, the canonical term 
$G_2=X$ gives rise to the contribution $\dot{\phi}^2/2$ 
to $\rho_{\phi}$. 
Existence of this term in Eq.~(\ref{back1}) is 
consistent with the relation (\ref{dphiH}).
Now, we search for scaling solutions obeying the 
relation (\ref{dphiH}) for an arbitrary constant $\alpha$.

For the realization of scaling solutions, we consider 
the case in which each term in $\rho_{\phi}$ and 
$P_{\phi}$ is proportional to $H^2$. 
Since $G_2$ is one of such terms, 
we require that $G_2 \propto H^2$. 
This relation translates to 
\be
\frac{\dot{G}_2}{H G_2}=2\frac{\dot{H}}{H^2}
=-3\left( 1+w_{\rm eff} \right)\,.
\label{dotG2}
\ee
On using Eq.~(\ref{dphiH}), the derivative 
$\dot{G}_2=G_{2,\phi}\dot{\phi}+G_{2,X} \dot{X}$ 
is expressed as 
\be
\dot{G}_2=H \left[ \alpha G_{2,\phi}-3(1+w_{\rm eff})
XG_{2,X} \right]\,.
\ee
Then, Eq.~(\ref{dotG2}) reduces to 
\be
X G_{2,X}-\frac{1}{\lambda}G_{2,\phi}-G_2=0\,,
\label{G2pa}
\ee
where 
\be
\lambda \equiv \frac{3(1+w_{\rm eff})}{\alpha}\,.
\ee
The partial differential Eq.~(\ref{G2pa}) is 
integrated to give 
\be
G_2(\phi,X)=X g_2 \left( Y \right)\,,
\label{G2lag}
\ee
where $g_2$ is an arbitrary function of 
\be
Y \equiv X e^{\lambda \phi}\,.
\label{Yd}
\ee

From Eq.~(\ref{dphiH}), the evolution of $\phi$
along the scaling solution is given by 
\be
\phi=\alpha \ln a+\phi_0\,,
\ee
where $\phi_0$ is a constant. 
For $w_{\rm eff}={\rm constant}$,  
the integrated solution to Eq.~(\ref{weff}) reads
\be
H=\frac{2}{3(1+w_{\rm eff})(t-t_0)}\,,\qquad 
a \propto (t-t_0)^{\frac{2}{3(1+w_{\rm eff})}}\,,
\label{Hubble}
\ee
where $t_0$ is a constant.  
Since $X \propto H^2 \propto (t-t_0)^{-2}$ 
and $e^{\lambda \phi} \propto a^{3(1+w_{\rm eff})} 
\propto (t-t_0)^2$ for scaling solutions, 
the quantity $Y=Xe^{\lambda \phi}$ does not vary 
in time.
Taking the $X$ derivative of Eq.~(\ref{G2lag}), 
it follows that 
$\dot{\phi}^2G_{2,X}=\dot{\phi}^2 (g_2+Yg_{2,Y}) 
\propto H^2$.
Hence the term $\dot{\phi}^2G_{2,X}$ in $\rho_{\phi}$ obeys 
the same scaling relation as $G_2$. 
We have thus shown that the quadratic Lagrangian 
$G_2(\phi,X)=X g_2 \left( Y \right)$, 
which was derived by the scaling property of one of 
the terms in $\rho_{\phi}$ and $P_{\phi}$, has the 
scaling solution. The result (\ref{G2lag}) coincides 
with that derived in Refs.~\cite{Piazza,Tsuji04} 
by assuming the nonvanishing constant $Q$.

For the cubic coupling $G_3$, the term 
$\dot{\phi}^2 G_{3,\phi}$ in $\rho_{\phi}$ and 
$P_{\phi}$ needs to be in proportion to 
$H^2 \propto \dot{\phi}^2$ for 
the existence of scaling solutions, so that 
$G_{3,\phi}={\rm constant}$. 
Taking the time derivative of this relation, 
it follows that 
\be
XG_{3,\phi X}-\frac{1}{\lambda}G_{3,\phi\phi}=0\,.
\ee
This is integrated to give
\be
G_3=g_3(Y)+h_3(X)\,,
\ee
where $g_3$ and $h_3$ are arbitrary functions of 
$Y$ and $X$, respectively. 
Since $G_{3,X}=e^{\lambda \phi}g_{3,Y}+h_{3,X}$, 
the terms $3H \dot{\phi^3}G_{3,X}$ and $-\dot{\phi}^2 \ddot{\phi}G_{3,X}$ in $\rho_{\phi}$ and 
$P_{\phi}$ are both proportional to 
$H \dot{\phi} Y \propto H^2$.
Hence the scaling relation is satisfied for any functional form of $g_3(Y)$. 

On the other hand, $h_3(X)$ gives rise to the terms 
proportional to $H^2X h_{3,X}$ in $\rho_{\phi}$ 
and $P_{\phi}$. In order to satisfy the scaling relation, 
we require that $X h_{3,X}=$ constant. This is 
integrated to give
\be
h_{3}(X)=c+d\ln X\,,
\ee
where $c$ and $d$ are constants. 
Then, the Lagrangian $-h_3(X)\square\phi$ is given by 
$(-c-d\ln Y+d\lambda\phi) \square \phi$, where we used 
the relation $\ln X=\ln Y-\lambda\phi$. 
The first term $-c\square \phi$ is just a total derivative. 
The second term $-(d\ln Y) \square \phi$ can be absorbed 
into $-g_3(Y) \square \phi$. 
The third term $d\lambda\phi\square\phi$ is equivalent to 
$2d\lambda X$ up to a boundary term, so it can be 
absorbed into the Lagrangian (\ref{G2lag}) by choosing $g_2(Y)=2d\lambda$. 
Then, the cubic interaction satisfying the scaling relation 
is simply expressed as 
\be
G_3(\phi, X)=g_3(Y)\,. 
\ee

{}From the above discussion, the scaling solution
with $\dot{\phi}/H={\rm constant}$ exists 
for the Lagrangian 
\be
L=Xg_2(Y)-g_3(Y) \square \phi\,,
\label{Lagsca}
\ee
where $Y$ is given by Eq.~(\ref{Yd}).
This result is valid not only for the nonvanishing $Q$ 
but also for $Q=0$.  
In  Ref.~\cite{Gomes1}, the authors obtained
the cubic Lagrangian of the form 
$g_3(Y)=a_1Y+a_2 Y^2$ by assuming a specific 
relation in the process of deriving the scaling Lagrangian.
As it is clear from the above discussion, any $Y$-dependent 
cubic coupling $g_3(Y)$, besides the Lagrangian 
$Xg_2(Y)$,  gives rise to scaling solutions. 
This is in agreement with the recent result 
of Ref.~\cite{Amen18}.

We also showed that the scaling Lagrangian 
can be expressed in the form 
$L=Xg_2(Y)-[g_3(Y)-d\lambda \phi] \square \phi$. 
If we choose $g_2(Y)=1-V_0/Y$ and $g_3(Y)=A\ln Y$, 
it accommodates the scaling model studied in Ref.~\cite{Alb}, 
i.e., the Lagrangian  $L=X-V_0e^{-\lambda \phi}
-A \ln (X e^{\bar{\lambda}\phi}) \square \phi$
with $\bar{\lambda}=\lambda (1-d/A)$.

\subsection{Field-dependent coupling $Q(\phi)$}
\label{scaderive2}

In this section, we derive the Lagrangian with scaling solutions 
for a field-dependent nonvanishing coupling $Q(\phi)$.
Employing  the scaling relation 
$\dot{\rho}_{\phi}/\rho_{\phi}=\dot{\rho}_m/\rho_m$ 
in Eqs.~(\ref{back4}) and (\ref{back3}), it follows that  
\be
\dot{\phi}=3\Omega_{\phi}
(w_m-w_{\phi}) \frac{H}{Q(\phi)}\,,
\label{dotphi}
\ee
and 
\be
\frac{\dot{\rho}_{\phi}}{H \rho_{\phi}}=
\frac{\dot{\rho}_{m}}{H \rho_{m}}=
-3(1+w_{\rm eff})\,.
\label{rhophim}
\ee
The field pressure $P_{\phi}=w_{\phi} \rho_{\phi}
\propto \rho_{\phi}$ also obeys the same relation 
as Eq.~(\ref{rhophim}), i.e., 
$\dot{P}_{\phi}/(H P_{\phi})=-3(1+w_{\rm eff})$.
This amounts to the scaling behavior 
$P_{\phi} \propto \rho_{\phi} \propto H^2$.

The Lagrangian (\ref{Lag}) contains the term $G_2$, which is
also present in $\rho_{\phi}$ and $P_{\phi}$. 
After integrating by parts the cubic Lagrangian $-G_3(\phi,X) \square \phi$
in the action (\ref{action}), the resulting Lagrangian 
contains the same terms which are present in $P_{\phi}$.
Thus, we search for the Lagrangian $L$ allowing for the same scaling 
property as $\rho_{\phi}$ and $P_{\phi}$, i.e., 
\be
\frac{\dot{L}}{HL}=
-3(1+w_{\rm eff})\,,
\label{Lcon}
\ee
or equivalently, $L \propto H^2$.
After deriving the Lagrangian satisfying the condition (\ref{Lcon}), 
we need to confirm whether each term in $\rho_{\phi}$ and $P_{\phi}$ 
obeys the scaling relation.

Since $L$ depends on $\phi, X$, and $\square \phi$, 
it follows that 
\be
\frac{\partial \ln L}{\partial \phi} 
\frac{\dot{\phi}}{H}+
\frac{\partial \ln L}{\partial \ln X} 
\frac{\dot{X}}{HX}+
\frac{\partial \ln L}{\partial \ln \square \phi} 
\frac{\dot{(\square \phi)}}{H \square \phi}
=-3\left( 1+w_{\rm eff} \right).
\label{pareq}
\ee
{}From Eq.~(\ref{dotphi}), the field derivative 
$X=\dot{\phi}^2/2$ is proportional to $H^2/Q^2(\phi)$. 
Then, we have
\be
\frac{\dot{X}}{HX}=-3 \left(1+w_{\rm eff} \right) 
\left( 1+\frac{2Q_{,\phi}}{\tilde{\lambda} Q^2} \right)\,,
\label{dotX}
\ee
where
\be
\tilde{\lambda} \equiv \frac{1+w_{\rm eff}}
{\Omega_{\phi}(w_m-w_{\phi})}\,.
\ee
Similarly, the term $\square \phi=-\ddot{\phi}-3H \dot{\phi}$ 
is expressed as 
\be
\square \phi=-\frac{3(1-w_{\rm eff}^2)P_{\phi}}
{2\tilde{\lambda} w_{\phi} \Omega_{\phi}Q}
+\frac{3(1+w_{\rm eff})^2 Q_{,\phi}P_{\phi}}
{\tilde{\lambda}^2 w_{\phi} \Omega_{\phi} Q^3}\,.
\label{sphi0}
\ee
Taking the time derivative of $\square \phi$, we find 
\be
\frac{\dot{(\square \phi)}}{H \square \phi}
=-3\left( 1+w_{\rm eff} \right) \left( 1+{\cal F} 
\right)\,,
\label{sphi}
\ee
where 
\be
{\cal F}=\frac{\tilde{\lambda} (w_{\rm eff}-1)Q^2 Q_{,\phi}
-2(w_{\rm eff}+1)(Q Q_{,\phi \phi}-3Q_{,\phi}^2)}
{\tilde{\lambda} Q^2 [\tilde{\lambda} (w_{\rm eff}-1)Q^2
+2(w_{\rm eff}+1)Q_{,\phi}]}\,.
\ee
Substituting Eqs.~(\ref{dotphi}), (\ref{dotX}), 
and (\ref{sphi}) into Eq.~(\ref{pareq}), we obtain 
\be
\frac{\partial \ln L}{\partial \phi} 
\frac{1}{\tilde{\lambda} Q}
-\frac{\partial \ln L}{\partial \ln X} 
\left( 1+\frac{2Q_{,\phi}}{\tilde{\lambda} Q^2} \right) 
-\frac{\partial \ln L}{\partial \ln \square \phi} 
\left( 1+{\cal F} \right)=-1\,.
\label{pareq2}
\ee
Plugging the Lagrangian (\ref{Lag}) into Eq.~(\ref{pareq2}), 
we find that the functions $G_2$ and $G_3$ need to 
satisfy the following relations
\ba
& & \left( 1+\frac{2Q_{,\phi}}{\tilde{\lambda} Q^2} \right) 
XG_{2,X}-\frac{1}{\tilde{\lambda} Q} G_{2,\phi}-G_2=0\,,\label{paeq1}\\
& &  \left( 1+\frac{2Q_{,\phi}}{\tilde{\lambda} Q^2} \right) 
XG_{3,X}+{\cal F} G_3 
-\frac{1}{\tilde{\lambda} Q} G_{3,\phi}=0\,.\label{paeq2}
\ea
{}From Eq.~(\ref{paeq1}), we obtain 
the following integrated solution 
\be
G_2(\phi,X)=Q^2(\phi)X\,\tilde{g}_2 
(\tilde{Y})\,,
\label{G2ge}
\ee
where $\tilde{g}_2$ is an arbitrary function of 
\be
\tilde{Y}=Q^2(\phi) X
e^{\tilde{\lambda} \psi}\,,
\ee
and 
\be
\psi=\int Q(\phi)\,d\phi\,.
\label{psi}
\ee
The integrated solution to Eq.~(\ref{paeq2}) is given by 
\be
G_3(\phi,X)=\frac{Q(\phi)\,\tilde{g}_3(\tilde{Y})}
{1+\mu\,Q_{,\phi}(\phi)/Q^2(\phi)}\,,
\label{G3ge}
\ee
where $\tilde{g}_3$ is an arbitrary function 
of $\tilde{Y}$, and 
\be
\mu=\frac{2(w_{\rm eff}+1)}
{\tilde{\lambda} (w_{\rm eff}-1)}\,.
\ee

Now, we will confirm whether each term in 
$\rho_{\phi}$ and $P_{\phi}$ 
is in proportion to $H^2$. 
On using the relations (\ref{Hubble}) and (\ref{dotphi})
in the scaling regime, the quantity (\ref{psi}) is given by
\be
\psi=\frac{2}{\tilde{\lambda}} \ln(t-t_0) +\psi_0\,,
\ee
where $\psi_0$ is a constant. 
Then, it follows that 
$\tilde{Y} \propto Q^2(\phi) \dot{\phi}^2 (t-t_0)^2 
\propto H^2 (t-t_0)^2={\rm constant}$. 
Hence the term $G_2$ in $\rho_{\phi}$ and $P_{\phi}$ 
has the dependence
$G_2 \propto Q^2(\phi) \dot{\phi}^2 \propto H^2$. 
Similarly, the term $\dot{\phi}^2 G_{2,X}$ in $\rho_{\phi}$ 
is proportional to $\dot{\phi}^2 Q^2(\phi)
(\tilde{g}_2+\tilde{Y} \tilde{g}_{2,\tilde{Y}}) \propto H^2$.

For the cubic coupling (\ref{G3ge}), its $X$ derivative
is given by 
$G_{3,X}=X^{-1}Q(\phi)\tilde{Y} \tilde{g}_{3,\tilde{Y}}/
(1+\mu Q_{,\phi}/Q^2)$. 
Then, only for $Q_{,\phi}/Q^2={\rm constant}$,  
the term $3H \dot{\phi}^3 G_{3,X}$ 
in $\rho_{\phi}$ is in proportion to $H^2$. 
Integration of this relation leads to 
\be
Q(\phi)=\frac{1}{c_1 \phi+c_2}\,,
\label{coupfo}
\ee
where $c_1$ and $c_2$ are constants. 
In this case, the other terms 
$-\dot{\phi}^2 G_{3,\phi}$ and $-\dot{\phi}^2 \ddot{\phi}G_{3,X}$ 
in $\rho_{\phi}$ and $P_{\phi}$ are also proportional to $H^2$.

In Ref.~\cite{Gomes1}, the coupling (\ref{coupfo}) was a priori 
assumed for simplicity to derive the Lagrangian with 
scaling solutions.
Here, we showed that the coupling is restricted to be 
of this form to realize the exact scaling properties 
of $\rho_{\phi}$ and $P_{\phi}$ associated with the 
cubic function $G_3(\phi,X)$.
Absorbing the constant in the denominator of 
Eq.~(\ref{G3ge}) into the definition of $\tilde{g}_3 (\tilde{Y})$, 
the cubic coupling with scaling solutions can be expressed as 
\be
G_3(\phi,X)=Q(\phi) \tilde{g}_3 (\tilde{Y})\,,
\label{G3f}
\ee
where 
\be
\tilde{Y}=Q(\phi)^{2-\tilde{\lambda}/c_1}X\,,
\ee
which is valid for $c_1 \neq 0$.

The coupling (\ref{coupfo}) includes the case of constant $Q$ 
(i.e., $c_1=0$).
In this case, we have $\psi=Q\phi$ and hence
the argument in the functions $\tilde{g}_2$ and $\tilde{g}_3$ 
reduces to $\tilde{Y}=Q^2X e^{\tilde{\lambda}Q \phi}$. 
Instead of $\tilde{\lambda}$, we define
\be
\lambda \equiv \tilde{\lambda}Q 
=\frac{Q(1+w_{\rm eff})}{\Omega_{\phi}(w_m-w_{\phi})}\,,
\ee
as well as $Y=Xe^{\lambda \phi}$, 
and absorb $Q$ into the definitions 
of $\tilde{g}_2$ and $\tilde{g}_3$. 
Then, the scaling Lagrangian 
can be written in the form 
\be
L=X g_2(Y)-g_3(Y) \square \phi\,,
\label{slag}
\ee
which coincides with Eq.~(\ref{Lagsca}). 

%%%%%%%%%%%%%%%%%%%%%%%%%%%%%%
\section{Fixed points for the dynamical system}
\label{Sec:fixed}
%%%%%%%%%%%%%%%%%%%%%%%%%%%%%%

In this section, we derive fixed points for the theories given by 
the Lagrangian (\ref{slag}) in presence of the constant coupling $Q$. 
Defining the dimensionless variables:
\be
x \equiv \frac{\dot{\phi}}{\sqrt{6}H}\,,\qquad 
y \equiv \frac{e^{-\lambda \phi/2}}{\sqrt{3}H}\,,
\label{xydef}
\ee
the quantity $Y=X e^{\lambda \phi}$ can be expressed as 
\be
Y=\frac{x^2}{y^2}\,.
\label{Ydef}
\ee
As we showed in Sec.~\ref{scaderive1}, $x$ and $Y$ are 
constant along the scaling solution, so $y$ is also 
constant from Eq.~(\ref{Ydef}). 

The dimensionless variables $x$ and $y$ obey the 
differential equations
\ba
x' &=& x \left( \epsilon_{\phi}- \epsilon_{h} 
\right)\,,\label{dxeq}\\
y' &=& -y \left( \frac{\sqrt{6}}{2}\lambda x 
+ \epsilon_{h}  \right)\,,\label{dyeq}
\ea
where a prime represents a derivative with respect to 
$N=\ln a$, and 
\be
\epsilon_{\phi} \equiv 
\frac{\ddot{\phi}}{H \dot{\phi}}\,,
\qquad
\epsilon_{h} \equiv \frac{\dot{H}}{H^2}\,.
\ee
The time derivatives $\ddot{\phi}$ and $\dot{H}$ 
are known by substituting $G_2=Xg_2(Y)$, $G_3=g_3(Y)$, 
and their $\phi, X$ derivatives into Eqs.~(\ref{ddotphi}) and (\ref{tH}).
{}From Eq.~(\ref{rhophi}), 
the field density parameter 
$\Omega_{\phi}=\rho_{\phi}/(3H^2)$ 
is given by 
\be
\Omega_{\phi}=x^2 \left( g_2+2Y g_{2,Y} \right)
-2Y g_{3,Y}x \left( \lambda x-\sqrt{6} \right)\,.
\label{Omephi}
\ee

The fixed points of the dynamical system 
(\ref{dxeq})-(\ref{dyeq}) can be derived by setting 
$x'=0$ and $y'=0$. Since the variables $x$ and $y$ 
are constant on fixed points, the quantity $Y=x^2/y^2$ 
and the functions $g_2(Y), g_3(Y)$ do not vary in time.  
The scaling solution discussed in Sec.~\ref{Sec:derivelag} 
obeys the following relations
\be
\epsilon_{\phi}=\epsilon_{h}
=-\frac{\sqrt{6}}{2}\lambda x_c\,.
\label{pare}
\ee
Here and in the following, we use the subscript ``$c$'' for 
the variables $x, y, Y$ in the case where they are 
associated with critical points of the dynamical system.
On using Eqs.~(\ref{ddotphi}) and (\ref{tH}), 
we obtain the following relation from Eq.~(\ref{pare}):
\be
\left[ 2(Q+\lambda)x_c-\sqrt{6} (1+w_m) \right]
\left[ \sqrt{6} \lambda x_c-3(1+g_2 x_c^2) \right]=0.
\label{scare}
\ee
There are two fixed points satisfying Eq.~(\ref{pare}). 
In the following, we will discuss each of them in turn.

\subsection{Point (a): Scaling solution}

One of the solutions to Eq.~(\ref{scare}) is given by 
\be
x_c=\frac{\sqrt{6} (1+w_m)}{2(Q+\lambda)}\,.
\ee
This corresponds to scaling solutions discussed in 
Sec.~\ref{Sec:derivelag}.
Indeed, the field density parameter $\Omega_{\phi}$ and 
the equation of state $w_{\phi}$ reduce, respectively, to 
\ba
\Omega_{\phi} 
&=& \frac{[2Q (Q+\lambda)+3(1+w_m)g_2 ](1+w_m)}
{2w_m (Q+\lambda)^2}\,,\label{Omephia} \\
w_{\phi}
&=& \frac{3w_m (1+w_m) g_2}
{3(1+w_m)g_2+2Q(Q+\lambda)}\,,
\ea
which are both constants. 
In the limit $Q \to 0$, we have $w_{\phi} \to w_m$
and hence $\rho_{\phi} \propto \rho_{m} 
\propto a^{-3(1+w_m)}$.
As we mentioned in Sec.~\ref{scaderive1}, the scaling solution 
exists not only for $Q \neq 0$ but also for $Q=0$.

We note that there are the following relations 
\ba
\hspace{-0.5cm}
& &
(1-w_m) g_2-2w_m Y_c g_{2,Y} \nonumber \\
\hspace{-0.5cm}
& &= -\frac{2Q(Q+\lambda)-6Y_c g_{3,Y} 
w_m (2Q+\lambda-w_m \lambda)}{3(1+w_m)},\label{aeq}\\
\hspace{-0.5cm}
& &
G_{2,X}=g_2+Y_cg_{2,Y}\,,\label{G2X}
\ea
which can be used to express $g_2$ and $g_{2,Y}$ 
in terms of $G_{2,X}$ and $g_{3,Y}$. 
Then, the density parameter (\ref{Omephia}) is
expressed as
\ba
\Omega_{\phi}&=& 
[Q(Q+\lambda)+3(1+w_m) G_{2,X} 
\nonumber \\
& &+3(2Q+\lambda-w_m \lambda) Y_cg_{3,Y}]
/(Q+\lambda)^2\,,
\label{Ophi2}
\ea
which explicitly shows the cubic-coupling contribution  
to $\Omega_{\phi}$. 
For $g_3=0$, Eq.~(\ref{Ophi2}) recovers the result 
derived in Ref.~\cite{Tsuji06}.
Taking the limit $Q \to 0$ in Eq.~(\ref{Ophi2}), 
we obtain
\be
\Omega_{\phi} \to \frac{3}{\lambda^2} 
\left[ (1+w_m) G_{2,X}+\lambda Y_c g_{3,Y} 
(1-w_m) \right]\,.
\ee
The quintessence with an exponential potential, which is 
given by the Lagrangian 
$G_2=X-V_0e^{-\lambda \phi}$, i.e.,  
$g_2(Y)=1-V_0/Y$, corresponds to the density parameter 
$\Omega_{\phi}=3(1+w_m)/\lambda^2$ \cite{CLW}. 

The effective equation of state $w_{\rm eff}=-1-2\epsilon_h/3$ 
reduces to
\be
w_{\rm eff}=\frac{w_m \lambda -Q}{Q+\lambda}\,.
\ee
If $Q=0$, then $w_{\rm eff}$ is equivalent to $w_m$. 
For $Q \neq 0$, the fixed point (a) 
can lead to the cosmic acceleration under 
the condition $(w_m \lambda-Q)/(Q+\lambda)<-1/3$. 
If we use this solution for the late-time cosmic 
acceleration with $w_{\rm eff}$ close to $-1$, 
the coupling $|Q|$ needs to be 
larger than the order $|\lambda|$. 
Since the CMB observations place the upper bound 
$|Q|<{\cal O}(0.1)$ \cite{Amenco2,Planckdark}, 
it is generally difficult 
to realize the scaling accelerated era characterized by 
$\Omega_{\phi} \simeq 0.7$ preceded by 
the scaling $\phi$MDE \cite{Amendola06}.
Hence we will not employ the fixed point (a) 
for the late-time cosmic acceleration.

In the limit $Q \to 0$, the fixed point (a) can be used for 
the scaling radiation and matter eras characterized by 
$w_{\rm eff}=w_{\phi}=w_m$. 
As we will show in Sec.~\ref{Sec:sta}, this scaling solution 
is typically a stable attractor for $\Omega_{\phi}<1$, 
so it does not exit from the scaling matter era.
If we want to realize the epoch of cosmic acceleration preceded 
by the scaling matter fixed point (a), 
the Lagrangian (\ref{slag}) needs to be 
modified in a suitable way at late time. 
In Refs.~\cite{Barreiro,Guo1,Guo2,Alb}, the authors took into 
account an additional scalar potential 
for achieving this purpose.

In this paper, we will pursue yet another possibility for realizing 
the cosmic acceleration preceded by the scaling $\phi$MDE 
without modifying the Lagrangian (\ref{slag}).

\subsection{Point (b): Scalar-field domination}

The other solution to Eq.~(\ref{scare}) corresponds to 
\be
g_2=\frac{\sqrt{6} \lambda x_c-3}{3x_c^2}\,.
\label{g2re}
\ee
{}From Eq.~(\ref{pare}), we also obtain
\be
g_{2,Y}=\frac{(\sqrt{6}-\lambda x_c)
(\sqrt{6}-6x_c Y_c g_{3,Y})}{6x_c^2 Y_c}\,.
\label{g2re2}
\ee
The field density parameter (\ref{Omephi}) 
reduces to  
\be
\Omega_{\phi}=1\,,
\ee
which means that the fixed point (b) satisfying the conditions 
(\ref{g2re}) and (\ref{g2re2}) is the scalar-field 
dominated point. 

On point (b), the effective equation of state and the field equation of 
state are equivalent to each other, such that 
\be
w_{\rm eff}=w_{\phi}=-1+\frac{\sqrt{6}}{3} 
\lambda x_c\,.
\label{weffwphi}
\ee
The cosmic acceleration occurs under the condition 
\be
\lambda x_c<\frac{\sqrt{6}}{3}\,.
\label{lamcon}
\ee
{}From Eqs.~(\ref{g2re}) and (\ref{g2re2}), we obtain 
the following relation
\be
\sqrt{6} G_{2,X}x_c=\lambda
+\left( \sqrt{6} \lambda x_c-6 \right) Y_c g_{3,Y}\,.
\ee
Then, Eq.~(\ref{weffwphi}) can be expressed as 
\be
w_{\rm eff}=w_{\phi}
=-1+\frac{\lambda^2+\lambda(\sqrt{6} \lambda x_c-6)
Y_cg_{3,Y}}
{3G_{2,X}}\,.
\label{wphib}
\ee
In the limit $\lambda \to 0$, we have $w_{\rm eff}=w_{\phi} \to -1$. 
Hence, for $\lambda$ close to 0, the fixed point (b) can be used 
for the late-time cosmic acceleration.
When $\lambda \neq 0$, the cubic coupling $g_3$ 
contributes to $w_{\rm eff}$ and $w_{\phi}$, 
whose property was shown in Ref.~\cite{Alb} for a specific choice 
of $g_3(Y)$. For given functions of $g_2(Y)$ and $g_3(Y)$, 
we can solve Eqs.~(\ref{g2re}) and (\ref{g2re2}) for $x_c$ and 
$Y_c$, so that $w_{\rm eff}$ and $w_{\phi}$ are known accordingly.

%%%%%%%%%%%%%%%%%%%%%%%%%
\begin{table*}[th!]
\small
\begin{center}
\begin{tabular}{|c|c|c|c|c|}
\hline
 & $x_c$  & $ \Omega_{ \phi}$ & $w_{\rm eff}$  
 & Stability\\ 
\hline\hline
(a) & $\frac{\sqrt{6} (1+w_m)}{2(Q+\lambda)}$ 
& Eq.~(\ref{Ophi2}) & 
$\frac{w_m \lambda -Q}{Q+\lambda}$  & Stable for 
$\frac{2Q+\lambda}{Q+\lambda}>0$, 
$\Omega_{\phi}<1$, $g_{2,Y}>0$
 \\ \hline 
(b) & known from Eqs.~(\ref{g2re})-(\ref{g2re2})  & 1 & 
$-1+\frac{\sqrt{6}}{3} \lambda x_{c}$ & 
Stable for $(Q+\lambda)x_{c}<\frac{\sqrt{6}}{2}$
\\ \hline
(c) & $\frac{\sqrt{6}[Q-3d_1(w_m-1)]}
{3(s+2d_1 Q)}$ & 
$\frac{2[ c_0 Q+d_1 \{3s+2Q
(6d_1-\lambda) \}]x_c}
{\sqrt{6}(s+2d_1 Q)}$ &  
$\frac{3w_m s-2Q[Q+3d_1 (1-2w_m)]}{3(s+2d_1Q)}$ & 
Saddle for $\frac{3c_0+2Q(Q+\lambda)}
{2c_0-4(Q+\lambda)d_1}>0$   \\ \hline
(d1) & $\frac{-\sqrt{6}d_1+ \sqrt{c_0+2d_1 (3d_1-\lambda)}}
{c_0-2d_1 \lambda}$ & 1 & 1 & 
Unstable for $|Qx_c|<\frac{\sqrt{6}}{2}$, 
$|\lambda x_c|<\sqrt{6}$  \\ \hline
(d2) & $\frac{-\sqrt{6}d_1- \sqrt{c_0+2d_1 (3d_1-\lambda)}}
{c_0-2d_1 \lambda}$ & 1 & 1 & 
Unstable for $|Qx_c|<\frac{\sqrt{6}}{2}$, 
$|\lambda x_c|<\sqrt{6}$  \\ 
\hline
\end{tabular}
\end{center}
\caption{\label{critp}
Critical points and corresponding values of $x_c$,  
$\Omega_{\phi}, w_{\rm eff}$ in the presence of 
a barotropic perfect fluid with the equation of state $w_m$. 
We also show the stability of fixed points for $w_m=0$. 
For given functions $g_2(Y)$ and $g_3(Y)$, 
the variables $Y_c$ and $y_c$ are known by solving 
Eq.~(\ref{aeq}) for point (a) and 
Eqs.~(\ref{g2re})-(\ref{g2re2}) for point (b).
The points (c), (d1), (d2), which satisfy $y_c=0$,
are present for the functions 
$g_2(Y)$ and $g_3(Y)$ given by Eqs.~(\ref{g2fun}) and (\ref{g3fun2}).}
\end{table*}
%%%%%%%%%%%%%%%%%%%%%%%

\subsection{Points (c) and (d1), (d2): Kinetic solutions}

We proceed to the second class of solutions to Eq.~(\ref{dyeq}), i.e., 
\be
y_c=0\,.
\ee
Let us consider the functions $g_2(Y)$ and 
$g_3(Y)$ containing the power-law functions of $Y$. 
The contributions arising from $g_2(Y)$ to 
$\rho_{\phi}$ and $P_{\phi}$ appear as the forms 
$X g_2$ and $XYg_{2,Y}$, while, for $g_3(Y)$, 
they arise as the combination $Yg_{3,Y}$.
In order to avoid the singular behavior 
at $Y =x^2/y^2\to \infty$, they are constrained 
to be of the forms
\ba
g_2(Y) &=& c_0+\sum_{n>0} c_n Y^{-n}\,,
\label{g2fun}\\
g_{3,Y}(Y) &=& \sum_{n \geq 1} d_n Y^{-n}\,,
\label{g3fun}
\ea
where $c_0, c_n, d_n$ and $n$ are constants. 
Integrating Eq.~(\ref{g3fun}) with respect to $Y$, 
it follows that 
\be
g_3(Y)=d_1 \ln Y+\sum_{n=2} \tilde{d}_n 
Y^{-n+1}\,,
\label{g3fun2}
\ee
where $\tilde{d}_n=d_n/(-n+1)$.  
Here, we omitted the integration constant 
$d_0$ in $g_3(Y)$, as it does not contribute to 
the cosmological dynamics.

Substituting the above expressions of $g_2(Y), g_3(Y)$ 
and their $Y$ derivatives into Eq.~(\ref{dxeq}) and taking 
the limit $Y \to \infty$, we obtain
\ba
x' &=& -[3(s+2d_1 Q)x-\sqrt{6} \{ Q-3d_1 (w_m-1) \}] 
\nonumber \\
& & \times [(c_0-2d_1\lambda)x^2+2\sqrt{6}d_1x-1] 
\nonumber \\
& &
\times[2c_0+4d_1(3d_1-\lambda)]^{-1}\,,
\label{dxd}
\ea
where 
\be
s \equiv \left( w_m-1 \right) \left( c_0-2d_1\lambda \right)\,.
\ee
The fixed point is determined by the coefficients 
$c_0$ and $d_1$ in Eqs.~(\ref{g2fun}) and (\ref{g3fun}).
{}From Eq.~(\ref{dxd}), there are the following two 
fixed points.

\begin{itemize}
\item Point (c): $\phi$MDE

One of the solutions to Eq.~(\ref{dxd}) 
is given by 
\be
x_c=\frac{\sqrt{6}[Q-3d_1(w_m-1)]}
{3(s+2d_1 Q)}\,.
\label{xpm}
\ee
On this fixed point (c), we have 
\ba
\hspace{-0.3cm}
\Omega_{\phi} &=& \frac{2[ c_0 Q+d_1 \{3s+2Q
(6d_1-\lambda) \}]x_c}
{\sqrt{6}(s+2d_1 Q)},\label{Omepp}\\
\hspace{-0.3cm}
w_{\phi} &=& \frac{Q[c_0+2d_1(6d_1 w_m-Q-\lambda)]
+3w_m d_1 s}{Q[c_0+2d_1 (6d_1-\lambda)]+3d_1 s},\\
\hspace{-0.3cm}
w_{\rm eff} &=&
\frac{3w_m s-2Q[Q+3d_1 (1-2w_m)]}{3(s+2d_1Q)}\,.
\label{weffp}
\ea
This is the scaling solution along which 
$\Omega_{\phi}, w_{\phi}, w_{\rm eff}$ are constant. 
In absence of the cubic coupling $d_1 \ln Y$ in $g_3(Y)$,  
we have $\Omega_{\phi}=w_{\rm eff}=2Q^2/(3c_0)$ and 
$w_{\phi}=1$ for $w_m=0$.
This is known as the $\phi$MDE \cite{Amenco1,Amenco2}, 
in which the dynamics of standard matter era is modified 
by the coupling $Q$.

Besides the constant $c_0$ in $g_2(Y)$, the function
$d_1 \ln Y$ in $g_3(Y)$ gives rise to contributions to 
the dynamics of $\phi$MDE.
This is a new $\phi$MDE solution corrected by the logarithmic 
cubic coupling $d_1 \ln Y$. 
The other terms on the right hand sides of 
Eqs.~(\ref{g2fun}) and (\ref{g3fun2}) do not 
modify the values of $\Omega_{\phi}, w_{\phi}, w_{\rm eff}$. 
Taking the limit $Q \to 0$ in Eqs.~(\ref{xpm}) and 
(\ref{Omepp})-(\ref{weffp}), we obtain  
$x_c=\sqrt{6}d_1/(2d_1 \lambda-c_0)$, 
$\Omega_{\phi}=6d_1^2/(2d_1\lambda-c_0)$, and 
$w_{\phi}=w_{\rm eff}=w_m$. 
This agrees with the fixed point (c) derived in Ref.~\cite{Alb} 
for the model $g_2(Y)=c_0+c_1/Y$ and $g_3(Y)=d_1 \ln Y$.
For this fixed point, the scalar sound speed 
squared is negative ($c_s^2=-1/3$). As we will see 
in Sec.~\ref{Sec:dark}, the $\phi$MDE with a nonvanishing 
coupling $Q$ can evade this problem.

\item Points (d1), (d2): Purely kinetic solutions

The other solutions to Eq.~(\ref{dxd}) are given by 
\be
x_c=\frac{-\sqrt{6}d_1 \pm \sqrt{c_0+2d_1 (3d_1-\lambda)}}
{c_0-2d_1 \lambda}\,,
\label{xd1d2}
\ee
where the plus and minus signs of $x_c$ correspond to  
the fixed points (d1) and (d2), respectively.
They are purely kinetic solutions, satisfying
\ba
\Omega_{\phi}=1\,,\qquad w_{\rm eff}=1\,,\qquad 
w_{\phi}=1\,,
\ea
which are relevant to neither radiation/matter eras 
nor the epoch of cosmic acceleration. 
\end{itemize}

In Table \ref{critp}, we summarize the fixed points and their properties.
In Sec.~\ref{Sec:sta}, we will study the stability of each point.

%%%%%%%%%%%%%%%%%%%%%%%%%%%%%%%%%
\section{Stability of fixed points}
\label{Sec:sta}
%%%%%%%%%%%%%%%%%%%%%%%%%%%%%%%%%

To study the stability of fixed points $(x_c, y_c)$ derived 
in Sec.~\ref{Sec:fixed}, we consider small homogeneous 
perturbations $\delta x$ and $\delta y$ around them, i.e., 
\be
x=x_c+\delta x\,,\qquad y=y_c+\delta y\,.
\ee
Then, the quantity $Y$ can be expressed as $Y=Y_c+\delta Y$, 
where the perturbation $\delta Y$ is expressed as
\be
\delta Y=2 \left( \frac{x_c}{y_c^2} \delta x-\frac{x_c^2}{y_c^3} 
\delta y \right)\,.
\ee
{}From Eqs.~(\ref{dxeq}) and (\ref{dyeq}), we obtain the linearized 
equations for $\delta x$ and $\delta y$ in the forms 
\ba
\left(
\begin{array}{c}
\delta x' \\
\delta y'
\end{array}
\right) = {\cal M} \left(
\begin{array}{c}
\delta x \\
\delta y
\end{array}
\right) \,, 
\label{uvdif}
\ea
where ${\cal M}$ is a $2 \times 2$ matrix given by 
\ba
{\cal M}=\left(
\begin{array}{c}
\frac{\partial x'}{\partial x}~~\frac{\partial x'}{\partial y} \\
\frac{\partial y'}{\partial x}~~\frac{\partial y'}{\partial y} 
\end{array}
\right)_{(x=x_c,y=y_c)}
 \equiv \left(
\begin{array}{c}
a_{11}~~a_{12} \\
a_{21}~~a_{22} 
\end{array}
\right).
\label{uvdif2}
\ea
The eigenvalues of ${\cal M}$ are 
\be
\mu_{\pm}=\alpha_{1}
\left( 1 \pm \sqrt{1-\alpha_2} \right)\,,
\label{deter}   
\ee
where 
\be
\alpha_1=\frac{a_{11}+a_{22}}{2}\,,\qquad 
\alpha_2=\frac{4(a_{11}a_{22}-a_{12}a_{21})}
{(a_{11}+a_{22})^2}\,.
\ee
If both $\mu_+$ and $\mu_-$ are negative or 
$\mu_{\pm}$ have negative real parts, then 
the fixed point is stable. 
When either $\mu_+$ or $\mu_-$ is positive, 
while the other is negative, 
it corresponds to a saddle point. 
If both $\mu_+$ and $\mu_-$ are positive, 
the fixed point is an unstable node.

In the following, we consider nonrelativistic 
matter characterized by 
\be
w_m=0\,,
\ee
as a background fluid.

\subsection{Point (a)}

The fixed point (a) corresponds to the scaling solution 
characterized by $x_c=\sqrt{6}/[2(Q+\lambda)]$.  
We first use Eqs.~(\ref{aeq}) and (\ref{G2X}) to 
eliminate $g_2$ from Eq.~(\ref{Ophi2}).
The term $g_{3,Y}$ can be 
expressed in terms of $\Omega_{\phi}$ and $g_{2,Y}$.
Then, we obtain $\alpha_1$ and $\alpha_2$ in 
Eq.~(\ref{deter}), as 
\ba
\alpha_1 
&=& -\frac{3(2Q+\lambda)}{4(Q+\lambda)}\,,\label{al1a}\\
\alpha_2
&=& 
\frac{24(1-\Omega_{\phi})g_{2,Y}}{q_s y_c^2 (2Q+\lambda)^2}\,,
\label{al2a}
\ea
where $q_s$ is defined by Eq.~(\ref{qs}). 
The term $g_{2,YY}$ appearing in $\alpha_2$ 
has been replaced with $q_s$.  
The condition for the absence of scalar ghosts 
in the small-scale limit corresponds to  
\be
q_s>0\,,
\label{qscon}
\ee
which is the same as that derived in 
Refs.~\cite{DT12,Kase18} for $Q=0$.

The stability of point (a) is ensured for $\alpha_1<0$ 
and $\alpha_2>0$.
Since  $\Omega_m>0$, the field density 
parameter should be in the range $\Omega_{\phi}<1$.
Then, the fixed point (a) is stable under the conditions 
\be
\frac{2Q+\lambda}{Q+\lambda}>0\,,\qquad 
\Omega_{\phi}<1\,,\qquad g_{2,Y}>0\,.
\label{sta}
\ee
The first and third conditions do not depend on $g_3$. 
{}From Eq.~(\ref{Ophi2}), the second condition translates to 
\be
\lambda^2>3 \left( G_{2,X}+\lambda Y_c g_{3,Y} \right)
+Q \left( 6 Y_c g_{3,Y}-\lambda \right)\,. 
\label{lambounda}
\ee
This means that $|\lambda|$ is generally bounded from below. 
If the right hand side of Eq.~(\ref{lambounda}) is of order 1, then 
$|\lambda| \gtrsim {\cal O}(1)$. 
In quintessence with the exponential potential $V(\phi)=V_0e^{-\lambda \phi}$, 
we have $g_2(Y)=1-V_0/Y$ with $V_0>0$ 
and hence $g_{2,Y}=V_0/Y^2>0$. 
The positivity of $g_{2,Y}$ also holds for the 
dilatonic ghost condensate model \cite{Piazza} given 
by the function $g_2(Y)=-1+cY$ with $c>0$. 

In the uncoupled case ($Q=0$), the first condition of 
Eq.~(\ref{sta}) is automatically satisfied.
The second condition $\Omega_{\phi}<1$ translates to 
$\lambda^2>3 \left( G_{2,X}+\lambda Y_c g_{3,Y} \right)$.  
This is consistent with the stability criterion derived in 
Ref.~\cite{Alb} for the model
$G_2=X-V_0e^{-\beta \phi}$ and $g_3=A \ln Y$.
Since $\beta=\lambda$ in the present case, 
the stability condition of point (a) reduces to 
$\lambda^2>3(1+A\lambda)$.

\subsection{Point (b)}

The scalar-field dominated point (b) satisfies the 
relations (\ref{g2re}) and (\ref{g2re2}). 
In this case, the eigenvalues $\mu_{\pm}$ yield
\ba
\mu_{+} &=& -3+\sqrt{6} \left( Q+\lambda \right)x_c\,,\\
\mu_{-} &=& -3+\frac{\sqrt{6}}{2}\lambda x_c\,,
\ea
which agree with those derived in Ref.~\cite{Tsuji06} in absence of 
the cubic coupling $g_3$. 
If the point (b) is responsible for the cosmic acceleration, 
we require that $w_{\rm eff}=-1+\sqrt{6}\lambda x_c/3<-1/3$, i.e., 
$\lambda x_c<\sqrt{6}/3$ and hence $\mu_{-}<-2$. 
Then, the stability of point (b) is ensured for 
\be
\left( Q+\lambda \right)x_c<\frac{\sqrt{6}}{2}\,.
\label{stab}
\ee
On using Eqs.~(\ref{g2re}) and (\ref{g2re2}) with 
$G_{2,X}=g_2+Y_c g_{2,Y}$, the variable $x_c$ 
can be expressed as
\be
x_c=\frac{\lambda-6Y_c g_{3,Y}}
{\sqrt{6} (G_{2,X}-\lambda Y_c g_{3,Y})}\,.
\ee
Then, the inequality (\ref{stab}) reads
\be
\frac{\lambda^2-3( G_{2,X}+\lambda Y_c g_{3,Y})
-Q \left( 6 Y_c g_{3,Y}-\lambda \right)}
{G_{2,X}-\lambda Y_c g_{3,Y}}<0\,.
\label{stab2}
\ee
If $G_{2,X}$ dominates over $\lambda Y_c g_{3,Y}$, i.e., 
\be
G_{2,X}>\lambda Y_c g_{3,Y}\,,
\label{G2Xc}
\ee
then the condition (\ref{stab2}) translates to 
\be
\lambda^2<3 \left( G_{2,X}+\lambda Y_c g_{3,Y} \right)
+Q \left( 6 Y_c g_{3,Y}-\lambda \right)\,,
\label{lamboundb}
\ee
which is exactly opposite to the stability condition 
(\ref{lambounda}) of point (a).
This means that, if the scalar-field dominated point (b)
is stable, the scaling solution (a) is not, 
and vice versa.

If the opposite inequality to Eq.~(\ref{G2Xc}) holds, 
then the term $\lambda(\sqrt{6} \lambda x_c-6)Y_cg_{3,Y}/(3G_{2,X})$ in Eq.~(\ref{wphib}) exceeds the order of 1. 
This leads to the large deviation of $w_{\phi}$ from $-1$, 
whose behavior is not observationally favored.
Hence it is natural to consider the  
inequality (\ref{G2Xc}), as 
it is the case for the specific model studied in Ref.~\cite{Alb}. 

For the application to dark energy studied later in Sec.~\ref{Sec:dark}, we resort to the point (b) as a late-time 
attractor with the cosmic acceleration. 
In this case, the scaling solution (a) is not stable, so 
it is irrelevant to the dark energy dynamics at late time. 

\subsection{Points (c) and (d1), (d2)}

The fixed point (c) corresponds to 
\be
x_c=-\frac{\sqrt{6}(Q+3d_1)}
{3[c_0-2(Q+\lambda)d_1]}\,,\qquad 
y_c=0\,,
\ee
during the matter dominance. 
In this case, the eigenvalues (\ref{deter}) reduce to 
\ba
\mu_{+}
&=&-\frac{3c_0-2Q^2-6(2Q+\lambda)d_1}
{2c_0-4(Q+\lambda)d_1}\,,\\
\mu_{-}
&=& \frac{3c_0+2Q(Q+\lambda)}
{2c_0-4(Q+\lambda)d_1}\,.
\ea
In terms of $\mu_{+}$, the field density parameter 
(\ref{Omepp}) can be expressed as 
\be
\Omega_{\phi}-1=\frac{2\mu_{+}}{3}
\left[ 1+\frac{(2\mu_{+}+3)d_1}{Q} \right]\,.
\ee
Provided that the ratio $|d_1/Q|$ is smaller than the order 1, 
we have $\mu_{+}<0$ for $\Omega_{\phi}<1$. 
If the condition 
\be
\mu_{-}=\frac{3c_0+2Q(Q+\lambda)}
{2c_0-4(Q+\lambda)d_1}>0\,,
\label{saddle}
\ee
is satisfied, the point (c) is a saddle. From the CMB 
observations, the coupling is constrained to be in the range 
$|Q| \lesssim {\cal O}(0.1)$ \cite{Planckdark}. 
Since we are considering the case $|d_1/Q|<{\cal O}(1)$, 
we have $|d_1| \lesssim {\cal O}(0.1)$.
If the scalar-field dominated point (b) corresponds to 
the late-time attractor, the quantity 
$\lambda$ is bounded as Eq.~(\ref{lamboundb}). 
For $c_0={\cal O}(1)$, $\lambda$ is typically 
smaller than the order 1, so that the condition 
(\ref{saddle}) is satisfied.
In this case, the $\phi$MDE point (c) is a saddle, 
which is followed by the stable point (b) 
with the cosmic acceleration.

For the points (d1) and (d2), the eigenvalues are 
given by 
\ba
\mu_{+}
&=& 3 \pm \sqrt{6}Q x_c\,,\\
\mu_{-}
&=& 3 \mp  \frac{\sqrt{6}}{2} \lambda x_c\,,
\ea
where the double signs are in the same orders as 
$x_c$ given in Eq.~(\ref{xd1d2}). 
If the conditions 
\be
\left| Qx_c \right|<\frac{\sqrt{6}}{2}\,,\qquad 
\left| \lambda x_c \right|<\sqrt{6}
\ee
are satisfied, these points are unstable nodes.

%%%%%%%%%%%%%%%%%%%%%%%%%%%%%%
\section{Application to dark energy}
\label{Sec:dark}
%%%%%%%%%%%%%%%%%%%%%%%%%%%%%%

Let us apply the theory given by the action (\ref{Lagsca}) 
to the dynamics of dark energy.
We assume that the scalar field $\phi$ is coupled to 
cold dark matter (density $\rho_c$ with vanishing 
pressure) with a constant coupling $Q$, such that 
\be
\dot{\rho}_{c}+3H \rho_{c}
=Q \rho_c \dot{\phi}\,.
\label{rhceq}
\ee
We take into account baryons (density $\rho_b$ with
vanishing pressure) and radiation (density $\rho_r$ and 
pressure $P_r=\rho_r/3$), which are both uncoupled to 
the field $\phi$. Then, the continuity equations 
are given, respectively, by 
\ba
& &
\dot{\rho}_b+3H \rho_b=0\,,\\
& &
\dot{\rho}_r+4H \rho_r=0\,.
\ea
\subsection{Cubic Horndeski theories with 
$\phi$MDE}

We are interested in the cosmological sequence of 
the $\phi$MDE followed by the cosmic acceleration 
driven by the fixed point (b). 
For this purpose, we consider 
the model given by the functions:
\ba
g_2(Y) &=& 1+\frac{c_1}{Y}\,,\label{g2con}\\
g_3(Y) &=& d_1 \ln Y-\frac{d_2}{Y}\,,\label{g3con}
\ea
which correspond to $n=1$ in (\ref{g2fun}) 
with $c_0=1$ and $n=2$ in (\ref{g3fun2}). 
The corresponding Lagrangian is given by 
\be
L=X+c_1 e^{-\lambda \phi}- \left( d_1 \ln Y 
-\frac{d_2}{Y} \right) \square \phi\,.
\label{lagcon}
\ee
For $d_2=0$, this reduces to the model studied in 
Ref.~\cite{Alb}. 
However, there are several important differences.
First of all, we take into account a nonvanishing coupling $Q$, 
which gives rise to the existence of $\phi$MDE. 
Secondly, the scaling fixed point (a) is not used for 
the early cosmological dynamics. 
If the point (a) is responsible for scaling radiation 
and matter eras, then the slope $\lambda$ of exponential 
potential $V(\phi)=-c_1 e^{-\lambda \phi}$ needs to obey
the condition (\ref{lambounda}). 
In this case, however, the stability condition (\ref{lamboundb}) 
of point (b) is not satisfied, so the system does not exit from 
the scaling solution (a) to the epoch of cosmic acceleration 
driven by point (b). 
The authors in Ref.~\cite{Alb} took into account another 
shallow exponential potential $V_2 e^{-\lambda_2 \phi}$  
for achieving this purpose.

In this paper, we do not modify the scaling Lagrangian 
(\ref{lagcon}) at late time. 
In this case, the slope $\lambda$ needs to satisfy the condition 
(\ref{lamboundb}), so that the solutions finally approach 
the stable fixed point (b) with cosmic acceleration.
We will study whether this fixed point (b) is preceded by the 
$\phi$MDE point (c) without having ghost and Laplacian 
instabilities. In the following, we study the case in which 
$|Q|$ and $|\lambda|$ are at most of the order 1.

In the small-scale limit, the scalar ghost is absent 
under the condition $q_s>0$, where $q_s$ is 
given by Eq.~(\ref{qs}). Expanding the action 
(\ref{action}) up to second order in scalar perturbations, 
one can show that the scalar propagation speed squared $c_s^2$ is 
of the same form as that derived in 
Refs.~\cite{DT12,Kase18}, i.e., 
\be
c_s^2=\frac{\xi_s}{q_s}\,,
\ee
where 
\ba
\xi_s
&=& 2[ (G_{3,XX}\ddot{\phi}+G_{3,X\phi}) 
\dot{\phi}^2+( 2\ddot{\phi}+4H \dot{\phi}) 
G_{3,X} \nonumber \\
& &+G_{2,X}-2G_{3,\phi} ]
-\dot{\phi}^4 G_{3,X}^2\,.
\ea
For the model (\ref{lagcon}), the quantities $q_s$ and 
$c_s^2$ reduce, respectively, to 
\ba
\hspace{-0.1cm}
q_s &=& 2+4d_1 \left( 3d_1 -\lambda \right) \nonumber \\
\hspace{-0.1cm}
& &
+\frac{4d_2y^2}{x^4} \left( 6d_1 x^2+3d_2 y^2 -\sqrt{6}x 
\right)\,,\label{qsex}\\
\hspace{-0.1cm}
c_s^2 &=& [ 6(1-2d_1^2-2d_1 \lambda)x^4
-24d_2 (d_1+\lambda)x^2 y^2 -12 d_2^2 y^4 \nonumber \\
\hspace{-0.1cm}
& & 
+4\sqrt{6} x \{ 2d_1 x^2+d_2 y^2 (2-\epsilon_{\phi})
\}]/(3q_sx^4)\,,\label{csex}
\ea
where
\begin{widetext} 
\ba
&&
\epsilon_{\phi}=\frac{1}{q_s}\Bigg[
6\sqrt6d_1(1-2 d_1 \lambda)x
-6(1-2d_1 \lambda -6 d_1^2)
-\frac{\sqrt6}{x}\big\{[c_1\lambda
-2d_2(3-12d_1\lambda -\lambda^2)] y^2
+d_1(6-3\omb-3\omc-4\omr)
\notag\\
&&\hspace{0.7cm}
+Q\omc\big\}
+\frac{24d_2(\lambda+3d_1)y^2}{x^2}
-\frac{\sqrt6d_2(12d_2 \lambda  y^2
+6-3\omb-3\omc-4\omr)y^2}{x^3}
+\frac{36d_2^2y^4}{x^4}
\Bigg]\,. 
\label{epphi}
\ea
\end{widetext} 
Here, we introduced the density parameters in the matter 
sector, as 
\be
\Omega_c=\frac{\rho_c}{3H^2}\,,\qquad 
\Omega_b=\frac{\rho_b}{3H^2}\,,\qquad
\Omega_r=\frac{\rho_r}{3H^2}\,.
\ee
For $d_1=0$ and $d_2=0$, we have $q_s=2$ and $c_s^2=1$, so 
there are neither ghost nor Laplacian instabilities. 
The cubic couplings $d_1$ and $d_2$ modify the values of 
$q_s$ and $c_s^2$.

Using the variables $x$ and $y$ defined in Eq.~(\ref{xydef}), 
the field density parameter $\Omega_{\phi}$ is expressed 
as $\Omega_{\phi}=\Omega_{G_2}+\Omega_{G_3}$, where 
\ba
\Omega_{G_2}
&=& x^2-c_1 y^2\,,\\ 
\Omega_{G_3}
&=& \frac{2}{x} \left( d_1 x^2+d_2 y^2 \right) 
\left( \sqrt{6}-\lambda x \right)\,.
\ea
Since $\rho_m=\rho_c+\rho_b+\rho_r$ in Eq.~(\ref{back1}), 
we obtain
\be
\Omega_c=1-\Omega_b-\Omega_r-\Omega_{G_2}
-\Omega_{G_3}\,.
\ee
The dark energy equation of state 
$w_{\phi}=P_{\phi}/\rho_{\phi}$ is given by 
\ba
\hspace{-0.3cm}
w_{\phi}
&=&[3(1-2d_1 \lambda )x^3
+3xy^2 (c_1-2d_2 \lambda) \nonumber \\
\hspace{-0.3cm}
& &
-2\sqrt{6}\,\epsilon_{\phi}(d_1 x^2+d_2 y^2)]
/[3(1-2d_1 \lambda )x^3
\nonumber \\
\hspace{-0.3cm}
& &-3xy^2 (c_1+2d_2 \lambda)
+6\sqrt{6} (d_1 x^2+d_2 y^2)
]\,.\label{wphimo}
\ea
The density parameters of baryons and radiation 
satisfy the differential equations
\ba
\Omega_b' &=& -\Omega_b \left( 3+2 \epsilon_h 
\right)\,,\\
\Omega_r' &=& -2\Omega_r  \left( 2+ \epsilon_h 
\right)\,,
\ea
where 
\begin{widetext} 
\ba
\hspace{-.5cm}
&&\epsilon_h=-\frac{1}{q_s}\Bigg[
6(1-2 d_1 \lambda )^2x^2
+12\sqrt6d_1(1-2d_1 \lambda )x
+6\lambda(c_1d_1-2d_2+6 d_1d_2 \lambda)y^2
+6d_1(6d_1+Q\omc)+(1-2 d_1 \lambda )(3\omb
\notag\\
\hspace{-.5cm}
&&\hspace{.75cm}
+3\omc+4\omr)
-\frac{24\sqrt6 d_1d_2\lambda y^2}{x}
+\frac{6d_2\{\lambda(c_1
+2d_2 \lambda )y^2+Q\omc\}y^2}{x^2}
-\frac{2\sqrt6 d_2(3\omb+3\omc+4\omr)y^2}{x^3}
-\frac{36d_2^2y^4}{x^4}\Bigg]\,.
\label{eph}
\ea
\end{widetext} 
The variables $x$ and $y$ obey the differential 
Eqs.~(\ref{dxeq}) and (\ref{dyeq}), where $\epsilon_{\phi}$ 
and $\epsilon_{h}$ are given, respectively, by 
Eqs.~(\ref{epphi}) and (\ref{eph}).

For the above dynamical system, the fixed point 
relevant to the radiation-dominated epoch is  
\ba
& &
x_c=-\frac{\sqrt{6}d_1}{1-2d_1 \lambda}\,,\qquad 
y_c=0\,,\qquad \Omega_b=0\,,\nonumber \\
& & \Omega_r=\frac{1+2d_1(3d_1-\lambda)}
{1-2d_1 \lambda}\,.
\label{radi}
\ea
On this fixed point with $d_1 \neq 0$, 
the quantities $q_s$ and $c_s^2$ 
reduce, respectively, to 
\ba
q_s &=& 2+4d_1 \left( 3d_1-\lambda \right)\,,\\
c_s^2 &=& -\frac{1}{3}\,.
\ea
Since $c_s^2<0$, the scalar perturbation is subject to 
Laplacian instabilities. 

For the theories with $d_2=0$, the scalar propagation 
speed squared is generally given by 
\be
c_s^2=\frac{3x(1-2d_1^2-2d_1 \lambda)+4\sqrt{6}d_1}
{3x(1+6d_1^2-2d_1 \lambda)}\,.
\ee
For $|d_1 \lambda|<{\cal O}(1)$, $|x|$ is the same order 
as $|d_1|$ around the radiation fixed point (\ref{radi}),  
in which case $c_s^2$ is negative. 
The only way of avoiding this instability problem 
is to consider the initial conditions satisfying 
$|d_1| \ll |x| \ll 1$, under which $c_s^2$ is close to 1. 
As long as the solutions do not approach the fixed 
point (\ref{radi}) during the radiation era, it is 
possible to avoid the Laplacian 
instability\footnote{If we use the fixed point (a) 
for realizing the scaling radiation era, there is a viable 
parameter space in which neither ghost nor Laplacian 
instabilities are present \cite{Alb}. 
In this case, unless the Lagrangian (\ref{slag}) is modified, 
the solutions do not exit from the scaling matter era 
to the epoch of cosmic acceleration.}. 
In such cases, however, the cubic coupling 
$g_3(Y)=d_1 \ln Y$ needs to be suppressed 
relative to $g_2(Y)=1+c_1/Y$
even in the early radiation era. 
Then, after the end of the radiation era, 
the effect of the cubic coupling on the scalar-field 
dynamics can be practically negligible.
Since the cosmological dynamics in such cases is 
indistinguishable from coupled quintessence 
with the exponential potential, we will not discuss 
the model $d_1 \neq 0$ any further.

\subsection{Model with $d_1=0$}

In the following, we study the model given by the 
Lagrangian (\ref{lagcon}) with
\be
d_1=0\,.
\ee
In this case, the fixed point associated with the 
radiation-dominated epoch is 
\be
x_c=0\,,\quad y_c=0\,,\quad \Omega_b=0\,,
\quad \Omega_r=1\,.
\label{pointra}
\ee
In realistic cosmology, the variables $x$ and $y$ do not 
exactly vanish during the radiation era. 
Substituting $\Omega_b=0$ and $\Omega_r=1$ into 
Eq.~(\ref{wphimo}) and $w_{\rm eff}=-1-2\epsilon_{h}/3$ 
and expanding them around $y_c=0$, it follows that 
\ba
w_{\phi} &=& 1+2 \left( c_1-Q d_2 \right) 
\frac{y^2}{x^2}+{\cal O}(y^4)\,,\label{wphira} \\
w_{\rm eff} &=& \frac{1}{3}+x^2
+\left[ c_1 x+2d_2 \left\{ \sqrt{6} -(Q+\lambda)x 
\right\} \right]\frac{y^2}{x} \nonumber \\
& &
+{\cal O}(y^4)\,.
\ea
As long as $|c_1 y^2/x^2| \ll 1$, $|d_2 y^2/x^2| \ll 1$, and 
$x^2 \ll 1$, it follows that  $w_{\phi} \simeq 1$ and 
$w_{\rm eff} \simeq 1/3$.

Similarly, the expansions of Eqs.~(\ref{qsex})-(\ref{csex}) 
around $y_c=0$ lead to 
\ba
\hspace{-0.5cm}
q_s &=& 2-4\sqrt{6} \frac{d_2y^2}{x^3}
+{\cal O}(y^4)\,,\\
\hspace{-0.5cm}
c_s^2 &=& 1+\frac{2}{3} \left[ 8\sqrt{6} 
-3\left( Q+2\lambda \right)x \right]
\frac{d_2y^2}{x^3}
+{\cal O}(y^4)\,.\label{csra}
\ea
Provided that $|d_2y^2/x^3| \ll 1$, 
we have $q_s \simeq 2$ and $c_s^2 \simeq 1$, so 
there are neither ghost nor Laplacian instabilities 
during the radiation dominance. 

The fixed point (c) corresponding to the $\phi$MDE is 
given by 
\be
x_c=-\frac{\sqrt{6}Q}{3}\,,\quad y_c=0\,,\quad
\Omega_c=1-\frac{2Q^2}{3}\,,\quad 
\Omega_r=0\,,
\label{pointc}
\ee
with 
\be
\Omega_{\phi}=\frac{2Q^2}{3}\,,\quad 
w_{\phi}=1\,,\quad w_{\rm eff}=\frac{2Q^2}{3}\,.
\label{OmephiM}
\ee
During the $\phi$MDE, we have 
\be
q_s=2\,,\qquad c_s^2=1\,,
\label{qsphi}
\ee
and hence neither ghost nor Laplacian 
instabilities are present.

For the scalar-field dominated point (b) relevant 
to the late-time cosmic acceleration, 
it is difficult to derive the analytic 
expressions of $x_c$ and $y_c$. 
For $d_2=0$, this fixed point corresponds to 
$x_c=\lambda/\sqrt{6}$ and 
$y_c=\sqrt{(\lambda^2-6)/(6c_1)}$.
Dealing with the cubic coupling $g_3(Y)=-d_2/Y$ as a 
correction to the leading-order solution derived 
for $d_2=0$, we obtain 
\ba
& &x_c=\frac{\lambda}{\sqrt{6}}
+\frac{(\lambda^2-6)^2}{\sqrt{6} c_1 \lambda^2}d_2
+{\cal O} (d_2^2)\,,\nonumber \\
& &y_c= \sqrt{\frac{\lambda^2-6}{6c_1}}
+{\cal O} (d_2^2)\,,
\nonumber \\
& &\Omega_b=0\,,\qquad \Omega_r=0\,, 
\label{pointb}
\ea
with $\Omega_{\phi}=1$.
For the validity of this solution, we  require that 
$|d_2 (\lambda^2-6)^2/(c_1 \lambda^3)| \ll 1$. 
Since we are considering the positive exponential 
potential ($c_1<0$), we require that $\lambda^2<6$.
We recall that the point (b) is stable 
under the condition (\ref{stab}). 
For $d_2=0$, this condition amounts to 
$(Q+\lambda)\lambda<3$.

On the fixed point (b) given by Eq.~(\ref{pointb}), 
the dark energy equation of state $w_{\phi}$ and 
the effective equation of state $w_{\rm eff}$ are 
\be
w_{\phi}=w_{\rm eff}=-1+\frac{\lambda^2}{3}
+\frac{(\lambda^2-6)^2}{3c_1 \lambda}d_2
+{\cal O} (d_2^2)\,.
\label{wphib2}
\ee
The quantities $q_s$ and $c_s^2$ can be estimated as
\ba
q_s &=& 2+\frac{24(6-\lambda^2)}{c_1 \lambda^3}
d_2+{\cal O}(d_2^2)\,,\label{qsb2} \\
c_s^2 &=& 1-\frac{2(6-\lambda^2)(10-\lambda^2)}
{c_1 \lambda^3}d_2+{\cal O}(d_2^2)\,.
\label{csb2}
\ea
Provided that the cubic coupling $d_2$ is suppressed 
relative to the leading-order terms  in Eqs.~(\ref{qsb2}) 
and (\ref{csb2}), there are neither ghost nor Laplacian 
instabilities.

%%%%%%%%%%%%%%%%%%%%%%%%%%%%%%
\begin{figure}[h]
\begin{center}
\includegraphics[height=3.3in,width=3.4in]{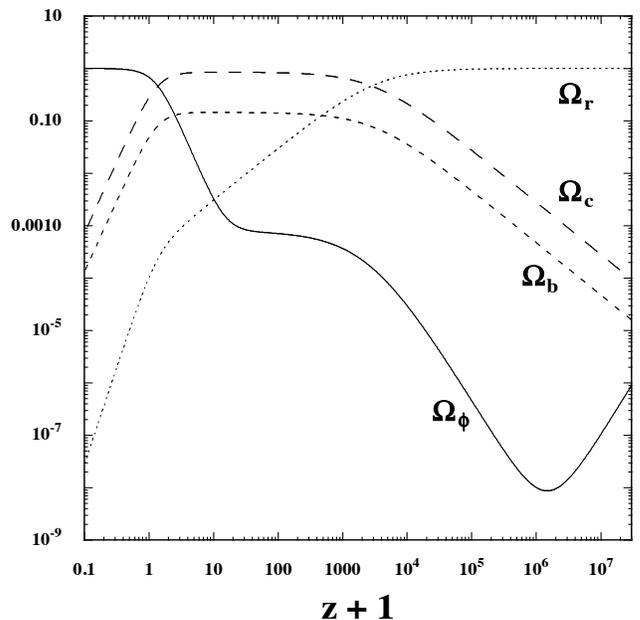}
\end{center}
\caption{\label{Omegafig}
Evolution of 
$\Omega_{\phi}, \Omega_c, \Omega_b, \Omega_r$ 
versus $z+1$ for $c_1=-1$, $d_1=0$, $d_2=10^{-3}$, 
$Q=0.04$, and $\lambda=-0.5$. The initial conditions are 
chosen to be $x=-10^{-3}$, $y=10^{-13}$, $\Omega_b=1.5 \times 10^{-5}$, 
and $\Omega_r=0.999895$ at the redshift $z=3.18 \times 10^7$. 
Today's field density parameter corresponds to $\Omega_{\phi}^{(0)}=0.681$.
}
\end{figure}
%%%%%%%%%%%%%%%%%%%%%%%%%%%%%%

%%%%%%%%%%%%%%%%%%%%%%%%%%%%%%
\begin{figure}[h]
\begin{center}
\includegraphics[height=3.3in,width=3.4in]{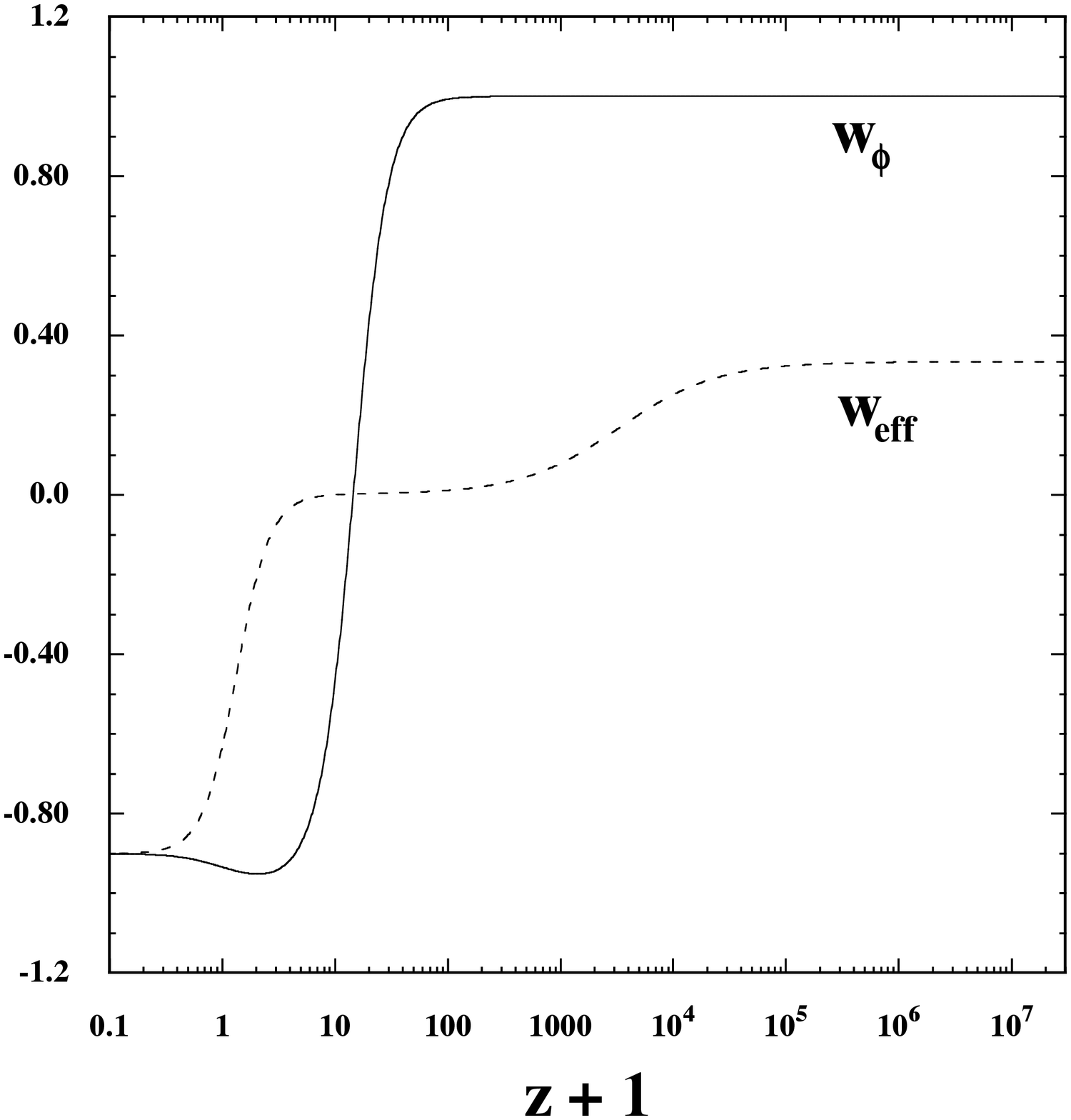}
\end{center}
\caption{\label{wfig}
Evolution of $w_{\phi}$ and $w_{\rm eff}$ for the same 
model parameters and initial conditions as those given 
in Fig.~\ref{Omegafig}.
}
\end{figure}
%%%%%%%%%%%%%%%%%%%%%%%%%%%%%%

%%%%%%%%%%%%%%%%%%%%%%%%%%%%%%
\begin{figure}[h]
\begin{center}
\includegraphics[height=3.3in,width=3.4in]{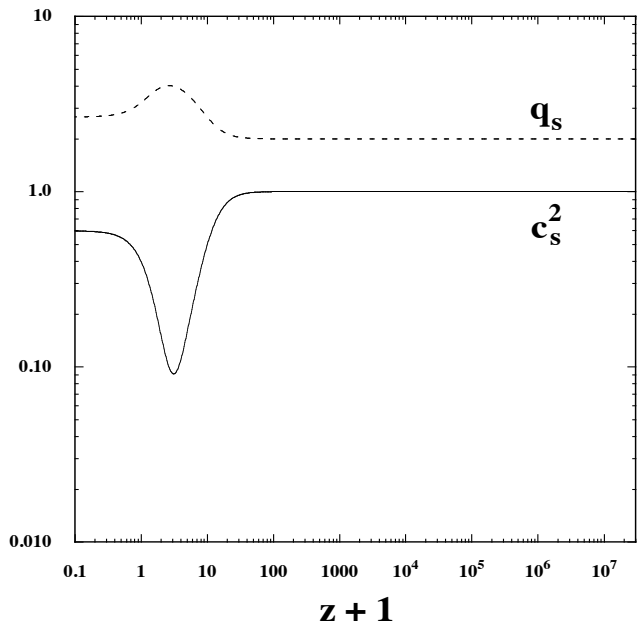}
\end{center}
\caption{\label{csfig}
Evolution of $q_s$ and $c_s^2$ for the same 
model parameters and initial conditions as those given 
in Fig.~\ref{Omegafig}.
}
\end{figure}
%%%%%%%%%%%%%%%%%%%%%%%%%%%%%%

In Fig.~\ref{Omegafig}, we plot the evolution of 
$\Omega_{\phi}, \Omega_c, \Omega_b, \Omega_r$ 
versus $z+1$ (where $z=1/a-1$ is the redshift) 
for the model parameters $c_1=-1$, $d_1=0$, $d_2=10^{-3}$, 
$Q=0.04$, and $\lambda=-0.5$. 
The corresponding variations of 
$w_{\phi}, w_{\rm eff}$ and $q_s, c_s^2$ 
are also shown in Figs.~\ref{wfig} and \ref{csfig}, 
respectively.  
The radiation fixed point (\ref{pointra}) is followed by the 
$\phi$MDE (\ref{pointc}) characterized by a nearly 
constant $\Omega_{\phi}$.
Since $Q=0.04$ in the numerical simulation of 
Fig.~\ref{Omegafig}, the $\phi$MDE corresponds to 
$x \simeq -0.03$ and $\Omega_{\phi} \simeq 0.001$.
The initial condition of $x$ in the radiation era is chosen as  
$x=-10^{-3}$, in which case $\Omega_{\phi}$ initially
decreases with the decrease of $|x|$. 
As we see in Fig.~\ref{Omegafig}, however, $\Omega_{\phi}$ 
starts to increase around the redshift $z=10^6$ toward the 
$\phi$MDE value $2Q^2/3$. 
In other words, even if $\Omega_{\phi}$ is initially as large as 
the background density parameters, the solutions approach 
the scaling $\phi$MDE with a nonnegligible 
dark energy density.
Thus, unlike the $\Lambda$CDM model, the field density 
does not need to be negligibly small relative to the 
background density even in the early 
radiation-dominated epoch.

For the scaling solution during the matter era,  the Planck team placed 
the bound $\Omega_{\phi}<0.02$ (95\,\%\,CL) 
around the redshift $z=50$ from the measurement of CMB 
temperature anisotropies \cite{Planckdark}. 
On using the value $\Omega_{\phi}=2Q^2/3$ 
during the $\phi$MDE, we obtain the upper limit 
$|Q|<0.17$. The coupling $Q$ used in Fig.~\ref{Omegafig} 
is consistent with this bound.

In Fig.~\ref{Omegafig}, the conditions  
$|c_1 y^2/x^2| \ll 1$, $|d_2 y^2/x^2| \ll 1$,  
$x^2 \ll 1$, and $|d_2y^2/x^3| \ll 1$ are satisfied during the 
radiation era, so that $w_{\phi} \simeq 1$, 
$w_{\rm eff} \simeq 1/3$, 
$q_s \simeq 2$, and $c_s^2 \simeq 1$ from 
Eqs.~(\ref{wphira})-(\ref{csra}). 
{}From Eqs.~(\ref{OmephiM})-(\ref{qsphi}), we have 
$w_{\phi}=1$, $w_{\rm eff} \simeq 0.001$, 
$q_s=2$, and $c_s^2=1$ during the $\phi$MDE.
Indeed, these properties can be confirmed in Figs.~\ref{wfig} and \ref{csfig}. 

As we observe in Fig.~\ref{Omegafig}, the field density parameter 
$\Omega_{\phi}$ starts to increase from the $\phi$MDE value  
$2Q^2/3$ toward the asymptotic value 1 around the redshift $z=20$. 
Since the $\phi$MDE under consideration corresponds to a saddle 
satisfying the condition (\ref{saddle}), the solution 
finally approaches 
the scalar-field dominated point (b). 
In this case, today's density parameters (at $z=0$) 
are $\Omega_{\phi}^{(0)}=0.681$, $\Omega_{c}^{(0)}=0.272$, 
$\Omega_{b}^{(0)}=0.047$, and $\Omega_{r}^{(0)}=1.0 \times 10^{-4}$.

In the numerical simulation of Fig.~\ref{Omegafig}, the asymptotic 
value of $x$ in the future is $x=-0.242$, so the stability 
condition (\ref{stab}) of point (b) is satisfied.
Up to the order of ${\cal O}(d_2)$, the approximate solutions 
(\ref{pointb}) and (\ref{wphib2}) give $x=-0.258$ 
and $w_{\phi}=-0.895$ for $\lambda=-0.5$, $c_1=-1$, and 
$d_2=10^{-3}$. They are slightly different from the numerical 
values $x=-0.242$ and $w_{\phi}=-0.901$. 
This difference is attributed to the fact that the contribution to $x$ 
arising from the cubic coupling $d_2$ is not very much smaller 
than the leading-order term $\lambda/\sqrt{6}$ 
for the model parameters used in Fig.~\ref{Omegafig}.

In Fig.~\ref{wfig}, we observe that $w_{\phi}$ starts to decrease 
from $1$ around the end of the $\phi$MDE and then temporally 
approaches the value close to $w_{\phi} \simeq -1$ 
at the redshift $z \approx 3$. 
Then, it grows toward the asymptotic value $w_{\phi}=-0.901$ of 
point (b). The evolution of $w_{\rm eff}$ is quite different from 
$w_{\phi}$ by today, but their asymptotic values are equivalent to 
each other. For the model parameters used 
in Fig.~\ref{wfig}, the Universe enters 
the stage of cosmic acceleration at the redshift $z<0.6$.

{}From Eqs.~(\ref{qsb2}) and (\ref{csb2}), the values of $q_s$ 
and $c_s^2$ on point (b) are in the ranges $q_s>2$ and $c_s^2<1$ 
for $d_2/(c_1 \lambda^3)>0$ with $\lambda^2 \lesssim 1$. 
This is the case for the numerical simulation of Fig.~\ref{csfig}, 
where $c_1<0$, $d_2>0$, and $\lambda<0$. 
In Fig.~\ref{csfig}, we find that $c_s^2$ decreases 
from 1 to the minimum value $0.091$ around $z=2.1$ 
and then it grows toward the asymptotic value $0.597$. 
Since both $q_s$ and $c_s^2$ are positive from the radiation 
era to the late-time accelerated attractor, there are neither ghost 
nor Laplacian instabilities of scalar perturbations. 
We note that the terms of order $d_2$ in Eqs.~(\ref{qsb2}) and (\ref{csb2})
give rise to contributions to $q_s$ and $c_s^2$ of order 1 for 
the model parameters used in Fig.~\ref{csfig}. 
This leads to the analytic values $q_s=3.104$ and 
$c_s^2=0.103$ on point (b), which do not exhibit good agreement with 
their asymptotic values seen in Fig.~\ref{csfig}.
The approximate formulas (\ref{qsb2}) and (\ref{csb2}) 
are valid only for $d_2/(c_1 \lambda^3) \ll 10^{-2}$. 

In the numerical simulations of Figs.~\ref{Omegafig}-\ref{csfig}, 
the density parameter arising from the cubic coupling is 
$\Omega_{G_3}=-0.018$ today, so it is by one order of 
magnitude smaller than the contribution $\Omega_{G_2}=0.699$. 
For increasing $d_2$, the contribution $\Omega_{G_3}$ to the 
total field density parameter $\Omega_{\phi}$ tends to be larger. 
At the same time, the minimum value of $c_s^2$ gets smaller and 
hence it can reach the instability region $c_s^2<0$. 
When $\lambda=-0.5$, this instability occurs for 
$d_2/(c_1 \lambda^3) \gtrsim 2 \times 10^{-2}$. 
For $\lambda=-{\cal O}(0.1)$ with 
positive $d_2/(c_1 \lambda^3)$, 
the criterion for avoiding the Laplacian instability is given by 
$d_2/(c_1 \lambda^3) \lesssim 10^{-2}$. 
If we consider $\lambda$ closer to 0, then the upper bound on 
$d_2/(c_1 \lambda^3)$ is generally loosened.
For negative $d_2/(c_1 \lambda^3)$, the sound speed 
squared (\ref{csb2}) becomes superluminal. 
Moreover, the upper bound on $|d_2/(c_1 \lambda^3)|$ for 
avoiding the instability at low redshifts tends to 
be severer relative to the case $d_2/(c_1 \lambda^3)>0$.

In summary, we have found an interesting scaling $\phi$MDE followed by 
the cosmic acceleration for the model given by the Lagrangian 
$L=X+c_1 e^{-\lambda \phi}+(d_2/Y) \square \phi$.
The cubic coupling $d_2$ modifies the conditions for the absence 
of ghost and Laplacian instabilities, but there are viable 
cosmological solutions without instabilities
like those shown in Figs.~\ref{Omegafig}-\ref{csfig}.
 
%%%%%%%%%%%%%%%%%%%%%%%%%%%%%%
\section{Conclusions}
\label{Sec:conclude}
%%%%%%%%%%%%%%%%%%%%%%%%%%%%%%

In this paper, we derived the most general Lagrangian in 
cubic-order Horndeski theories allowing for the existence of 
cosmological scaling solutions with 
the field-dependent coupling $Q(\phi)$. 
The functions $G_2$ and $G_3$ are restricted to be
Eqs.~(\ref{G2ge}) and (\ref{G3f}), respectively, to 
realize the scaling behavior 
$P_{\phi} \propto \rho_{\phi} \propto H^2$.
We showed that, in the presence of the cubic Lagrangian 
$-G_3(\phi,X) \square \phi$, the coupling $Q(\phi)$ is 
constrained to be of the form $Q(\phi)=1/(c_1 \phi+c_2)$. 
Unlike the analysis of Ref.~\cite{Gomes1} in which the same 
coupling is a priori assumed, we derived its form 
from the scaling relation of each term 
in the background equations of motion.

For the constant coupling $Q$, which includes $Q=0$ as 
a special case, the Lagrangian reduces to the form (\ref{slag}).
Unlike Ref.~\cite{Gomes1} the cubic coupling is not restricted to 
be of the form $G_3=a_1Y+a_2Y^2$, but $G_3$ is an arbitrary  
function $g_3(Y)$ with respect to $Y=Xe^{\lambda \phi}$. 
This property is consistent with the recent finding of Ref.~\cite{Amen18}.

In Sec.~\ref{Sec:fixed}, we obtained the fixed points of the dynamical 
system with a matter perfect fluid for the theories 
given by the Lagrangian (\ref{slag}) with constant $Q$.
Without specifying any functional forms of $g_2(Y)$ and $g_3(Y)$, 
we derived the two fixed points (a) and (b) corresponding to the 
scaling solution and the scalar-field domination, respectively. 
For the functions given by Eqs.~(\ref{g2fun}) and (\ref{g3fun2}), 
we showed the existence of the $\phi$MDE fixed point (c)  
besides the purely kinetic solutions (d1) and (d2). 
We note that the $\phi$MDE does not exist for the cubic coupling 
of the form $g_3(Y)=a_1Y+a_2Y^2$ derived in Ref.~\cite{Gomes1}.

In Sec.~\ref{Sec:sta}, we studied the stability of fixed points by considering 
homogenous perturbations around them. We showed that, if the scalar-field 
dominated point (b) is a stable attractor, the scaling solution (a) is not stable, 
and vice versa. The $\phi$MDE fixed point (c) is a saddle under the 
condition (\ref{saddle}). The $\phi$MDE, which exists in the presence 
of the nonvanishing coupling $Q$, can be followed by the epoch of cosmic 
acceleration driven by point (b). Since the coupling is constrained to be 
in the range $|Q|<{\cal O}(0.1)$ from CMB observations, the point (a) is difficult to be 
used as a scaling accelerated attractor with $\Omega_{\phi} \simeq 0.7$. 
The point (a) can be applied to the scaling radiation and matter eras, but 
in this case the Lagrangian (\ref{slag}) needs to be modified to 
exit from the scaling regime to the epoch of cosmic acceleration 
(as studied in Ref.~\cite{Alb}). 
In this paper, we employ the scaling $\phi$MDE point (c)
instead of the scaling solution (a) for the matter era, 
without modifying the scaling Lagrangian  (\ref{slag}).

In Sec.~\ref{Sec:dark}, we proposed a concrete model of dark energy 
given by the Lagrangian (\ref{lagcon}) with the $\phi$MDE followed 
by the scalar-field dominated point (b). 
For the model with $d_1 \neq 0$, there exists the radiation-dominated 
fixed point (\ref{radi}), on which the scalar sound speed squared 
is negative ($c_s^2=-1/3$). To avoid the Laplacian instability, 
the variable $x$ initially needs to be in the range $|d_1| \ll |x| \ll 1$. 
In such cases it is possible to evade the instability problem, 
but the effect of the cubic coupling on the scalar-field dynamics is 
practically negligible after the end of the radiation era.

On the other hand, the model with $d_1=0$ gives rise to interesting
cosmological solutions where the cubic coupling $g_3(Y)=-d_2/Y$ 
provides an important contribution to the late-time dynamics of 
dark energy.  As we observe in Fig.~\ref{Omegafig}, there exists the 
$\phi$MDE followed by the scalar-field dominated point (b). 
Moreover, the field density parameter $\Omega_{\phi}$ does 
not need to be negligibly small in the early radiation era, 
so it alleviates the small energy-scale problem of the 
$\Lambda$CDM model. After the $\phi$MDE 
the dark energy equation of state 
$w_{\phi}$ starts to decrease from 1, 
reaches a minimum close to $-1$ at low redshifts, 
and then finally approaches the value (\ref{wphib2}) of  
point (b), see Fig.~\ref{wfig}. As we showed in 
Fig.~\ref{csfig}, there exists the viable model parameter space 
in which the conditions for the absence of ghost and Laplacian 
instabilities are satisfied even in the presence of the cubic coupling.

It will be of interest to place observational constraints on the model 
proposed in Sec.~\ref{Sec:dark} by extending the EFTCAMB code 
developed in \cite{EFT1,EFT2}.
For coupled quintessence without the cubic coupling, 
the likelihood analysis based on the Planck CMB data combined with 
the data of baryon acoustic oscillations and weak lensing 
showed that  there is a peak around 
$|Q| = 0.04$ for the marginalized posterior distribution 
of $Q$ \cite{Petto,Planckdark}. 
It remains to be seen whether this property also persists
when the cubic coupling is present.

%%%%%%%%%%%%%%%%%%%%%%%%%%%%%%
\begin{acknowledgements}
We thank Luca Amendola, Guillem Dom\`enech, and 
Adalto R. Gomes for useful correspondence.
The research of NF and NJN is supported by Funda\c{c}\~{a}o para a  
Ci\^{e}ncia e a Tecnologia (FCT) through national funds  (UID/FIS/04434/2013), 
by FEDER through COMPETE2020  (POCI-01-0145-FEDER-007672) and  
by FCT project ``DarkRipple- Spacetime ripples in the dark gravitational Universe'' 
with reference PTDC/FIS- OUT/29048/2017.
NJN  is also supported by an FCT Research contract, with reference IF/00852/2015. 
RK is supported by the Grant-in-Aid for Young Scientists B 
of the JSPS No.\,17K14297. 
ST is supported by the Grant-in-Aid 
for Scientific Research Fund of the JSPS No.~16K05359 
and MEXT KAKENHI Grant-in-Aid for 
Scientific Research on Innovative Areas ``Cosmic Acceleration'' (No.\,15H05890). 
\end{acknowledgements}
%%%%%%%%%%%%%%%%%%%%%%%%%%%%%%

%%%%%%%%%%%%%%%%%

\end{document}